\documentclass[final]{ias2}

\usepackage{graphicx}
\usepackage{multirow}
\usepackage{array}

\usepackage{hyperref}

\newcommand{\bea}{\begin{eqnarray}}
\newcommand{\eea}{\end{eqnarray}}
\newcommand{\bes}{\begin{subequations}}
\newcommand{\ees}{\end{subequations}}
\newcommand{\ds}{\displaystyle}

\begin{document}

\markboth{Novel energy sharing collisions of multicomponent solitons}{T Kanna, K Sakkaravarthi and M Vijayajayanthi}

\title{Novel energy sharing collisions of multicomponent solitons}

\author[bhc]{T Kanna}\email{Corresponding author: kanna\_phy@bhc.edu.in, kanna.phy@gmail.com}
\author[bhc]{K Sakkaravarthi}\email{ksakkaravarthi@gmail.com}
\author[anna]{M Vijayajayanthi}\email{vijayajayanthi.cnld@gmail.com}
\address[bhc]{Post Graduate and Research Department of Physics, Bishop Heber College, Tiruchirappalli -- 620 017, Tamil Nadu, India}
\address[anna]{Department of Physics, Anna University, Chennai -- 600 025, Tamil Nadu, India}
\begin{abstract}
In this paper, we discus the fascinating energy sharing collisions of multicomponent solitons in certain incoherently coupled and coherently coupled nonlinear Schr\"odinger type equations arising in the context of nonlinear optics.
\end{abstract}



\maketitle

\section{Introduction}
Solitons are fascinating nonlinear entities with huge potential for technological applications due to their remarkable collision properties. Following the pioneering numerical work of Zabusky and Kruskal \cite{Kruskal-prl} on soliton collisions, there has been a number of research papers on soliton interaction and still it remains as a frontier topic of research. It is well known that solitons are solitary waves that asymptotically preserve their amplitude and speed during its collision with other solitary wave except for a phase-shift \cite{Ablo-ist-book}.

Multicomponent solitons (MSs) are intriguing nonlinear objects in which a given soliton is split among several components. These solitons are also known as vector solitons or multi-color solitons. From a mathematical perspective these MSs arise as solutions of certain multicomponent integrable nonlinear partial differential equations. In integrable systems, such MSs have same central position and travel with same velocity. These MSs appear in wide range of physical systems that include nonlinear optics \cite{Akm-book,Kiv-book}, plasma physics \cite{Akm-book}, water waves \cite{Shukla-prl}, bio-physics \cite{Scot} and Bose-Einstein condensates \cite{Frantz-bec}. Here our focus will be on particular MSs arising in the context of nonlinear optics.

In the context of nonlinear optics, the MSs arise as solutions of integrable multiple coupled nonlinear Schr\"odinger type equations which describe the dynamics of simultaneous propagation of multiple waveguide modes in Kerr like media \cite{Kiv-book}. Such multicomponent systems show interesting propagation and collision dynamics as a result of various nonlinear effects.  When two or more optical modes co-propagate inside a fiber, they can interact with each other through the fiber nonlinearity. In general, such interactions are governed by coupled nonlinear Schr\"odinger (CNLS) family of equations. Based on the presence and absence of coherent (phase-dependent) nonlinearities, these CNLS equations can be classified into two classes, namely coherently coupled nonlinear Schr\"odinger (CCNLS) equations and incoherently coupled nonlinear Schr\"odinger (ICNLS) equations, respectively. A physically interesting set of ICNLS equations arising in nonlinear optics is
\bea
&&i q_{j,z}+ q_{j,tt}+2 \left(|q_1|^2+\sigma |q_2|^2\right) q_j =0, \quad ~~j=1,2,
\label{cnls}
\eea
in which the nonlinear couplings are due to self-phase modulation (SPM) and cross-phase modulation (XPM) and depend only on the local intensities of the co-propagating fields, but insensitive to their phases \cite{Kiv-book}. For $\sigma=1$, Eqn. (\ref{cnls}) reduces to the integrable Manakov system with $q_j$, $j=1,2$, being the envelope of the $j$th mode, $z$ and $t$ represent the normalized distance along the fiber and the retarded time, respectively and describes an intense electromagnetic pulse propagation in birefringent fiber \cite{Manakov}. For $\sigma=-1$, system (\ref{cnls}) becomes as the mixed-ICNLS system. Lazarides and Tsironis \cite{lazar} have obtained this mixed ICNLS system as governing equations for electromagnetic pulse propagation in isotropic and homogeneous nonlinear left handed materials, by taking the effective permittivity and effective permeability to be intensity dependent and following a reductive perturbational approach. Here $q_1$ and $q_2$ are the electric and magnetic field components of the electromagnetic pulse, respectively, the subscripts $z$ and $t$ denote the partial derivatives with respect to normalized distance and retarded time respectively. Mixed-ICNLS system (\ref{cnls}) can also be obtained as the modified Hubbard model (Lindner-Fedyanin system) in the long-wavelength approximation by taking the electron-phonon interaction into account \cite{lindner}. These Manakov and mixed-ICNLS systems find important applications in optical communication and in artificial metamaterials. They have been intensively studied in literature \cite{Manakov,lazar,lindner,RK1997pre,Ablo-ist-book,Ablo-book2,Kannapramana,Kanna2001prl,Kanna2003pre,Kanna2006pre,recent,Kanna2008pra,Kanna2009epjst}. Also, the integrable multicomponent generalization of ICNLS system (\ref{cnls}) can be written as
\bea
&&i q_{j,z}+ q_{j,tt}+2 \left(\sum_{j=1}^m \sigma_j |q_j|^2\right) q_j =0, \quad ~~j=1,2,3,...,m,
\label{mcnls}
\eea
where $\sigma_j=\pm 1$ represent the nature of nonlinear coupling, which is of either focusing (Manakov) type for $\sigma_j=1$ and defocusing type for $\sigma_j=-1$ or mixed type ($\sigma_j=1$ for $j=1,2,...,p$ and $\sigma_j=-1$ for $j=p+1,p+2,...,m$). The above system admits bright soliton solutions for the Manakov (focusing) case and it supports both bright and bright-dark soliton solutions for the mixed-ICNLS case \cite{RK1997pre,Ablo-book2,Kannapramana,Kanna2001prl,Kanna2003pre,Kanna2006pre,Kanna2008pra,Kanna2009epjst}. Particularly, the bright multi-soliton solutions of the multicomponent generalization of Manakov system have been obtained by Kanna {\it et al.} using the Hirota bilinearization method and a detailed investigation on the soliton collisions, such as energy sharing collision and elastic type interactions, have been explored \cite{Kanna2001prl,Kanna2003pre,Kanna2006pre,Kanna2009epjst}. In this paper, we will review the results of two component systems only.

In general cases, like pico-second pulse propagation in non-ideal low birefringent multimode fibers or beam propagation in weakly anisotropic Kerr type nonlinear media, the coherent effects due to the interaction of co-propagating fields should also be considered \cite{Kiv-book,crosi}. The propagation of coherently coupled orthogonally polarized waveguide modes in Kerr type nonlinear medium is governed by the following $2$-component coherently coupled nonlinear Schr\"odinger (CCNLS) type equations \cite{Kiv-book,crosi,Park};
\bes\bea
i q_{1,z}+ \delta q_{1,tt}-\mu q_{1}+ (|q_{1}|^2+\sigma |q_{2}|^2)q_{1}+\lambda q_2^2 q_{1}^*=0,\\
i q_{2,z}+\delta q_{2,tt}+\mu q_{2}+ (\sigma |q_{1}|^2+|q_{2}|^2)q_2+\lambda q_{1}^2 q_{2}^*=0,
\eea \label{e1} \ees
where $q_1$ and $q_2$ are slowly varying complex amplitudes in each polarization mode, $z$ and $t$ are the propagation direction and transverse direction, respectively, $\mu$ is the degree of birefringence and $\delta$ is the group velocity dispersion.

In the above equations, the nonlinearities arise from the SPM ($|q_k|^2q_j$, $j=k=1,2$), XPM ($|q_k|^2q_j$, $j,k=1,2,~k\neq j$) and four-wave mixing process (FWM: $q_k^2q_j^*$, $j,k=1,2,~k\neq j$), among which the first two are phase-independent while the third one is phase-dependent (coherent) nonlinearity. Also, Eq. (\ref{e1}) is non-integrable. However for specific choices of system parameters ($\delta,~\mu,~\sigma$ and $\lambda$) it becomes integrable and exist in different physical situation \cite{Park}. The corresponding integrable CCNLS system is
\bes \bea
i q_{1,z}+ q_{1,tt}+ \gamma (|q_{1}|^2+2 |q_{2}|^2)q_{1}-\gamma q_2^2 q_{1}^*=0,\\
i q_{2,z}+ q_{2,tt}+ \gamma (2 |q_{1}|^2+|q_{2}|^2)q_2-\gamma q_{1}^2 q_{2}^*=0.
\eea \label{2cceqn} \ees
Hereonwards we refer to the above system as 2-CCNLS system. The above system governs the dynamics of pulse propagation in nonlinear gyrotropic media \cite{gyro} as well as in an isotropic nonlinear Kerr medium for particular choices of third order susceptibilities. System (\ref{2cceqn}) is shown to be integrable by Painlev\'e analysis and soliton solutions were obtained as the linear superposition of two nonlinear Schr\"odinger (NLS) solitons \cite{Park}.

An integrable multicomponent generalization of the above 2-CCNLS system (\ref{2cceqn}) is,
\bea
\hspace{-1.5cm}i q_{j,z}+ q_{j,tt}+\gamma \left(|q_j|^2+2 \sum_{l=1,l\neq j}^2 |q_l|^2\right) q_j - {\gamma} \sum_{l=1,l\neq j}^2 q_l^2 q_j^*=0, \quad j=1,2,3,...,m.~~
\label{cceqn}
\eea
In Ref. \cite{Kanna2010jpa}, Kanna {\it et al}., have also investigated bright soliton dynamics in another type of 2-CCNLS system similar to the two-component version of (\ref{2cceqn}). This 2-CCNLS system shows novel energy switching collision of bright solitons as will be discussed below. Additionally, there exists another integrable 2-CCNLS system with nonlinearities having opposite signs in the two components, for which the soliton solutions and bound states are constructed in Ref. \cite{Kanna2013jmp}. As the solitons in this system undergo standard elastic collision we do not discuss this system in this review.

The soliton solutions of multicomponent Manakov and mixed-ICNLS systems (\ref{mcnls}) were obtained using the Hirota's bilinearization method \cite{Hirota-book} by transforming the nonlinear equations (\ref{mcnls}) into the bilinear form and by recursively solving the resulting a set of equations in a standard way. On the other hand, for $m$-CCNLS system (\ref{cceqn}) we have to apply non-standard bilinearization procedure. A standard bilinearization procedure will result in  a greater number of bilinear equations  than the number of bilinearising variables, which results in soliton solutions with less number of arbitrary parameters. In order to get more general soliton solutions we introduce an auxiliary function during the bilinearization of the $m$-CCNLS system which gives equal number of bilinear equations and variables \cite{Gilson,Hirota-book,Kanna2010jpa,Kanna2011jpa,Kanna2013jmp}. To be more clear with the presentation, we give below the  bilinear equations for the $m$-CCNLS system (\ref{cceqn})
\bes\label{beq}\bea
(iD_z+D_t^2)(g^{(j)}\cdot f) &=& \gamma s g^{(j)*}, \quad\quad\quad\quad j=1,2,...,m,\\
D_t^2(f \cdot f) &=& 2 \gamma \sum_{j=1}^m |g^{(j)}|^2, \\
s\cdot f&=&\sum_{j=1}^m (g^{(j)})^2.
\eea  \label{3beq} \ees
obtained by using the rational transformation $q_j=\frac{g^{(j)}}{f}$, $j=1,2,...,m$, with the introduction of an auxiliary function $s$. Here $g^{(j)}$ ($f$) is complex (real) function of $z$ and $t$, $D_z$ and $D_t$ represent  the Hirota's differential operators \cite{Hirota-book} and $*$ indicates the complex conjugation. The exact forms of $g^{(j)}$, $f$, and $s$, which result  in  the soliton solutions of (\ref{cceqn}), are obtained by suitably expressing them as power series and recursively solving the resulting set of equations at various powers of expansion parameter from the bilinear equations (\ref{3beq}). One can refer to \cite{Kanna2011jpa} for more details regarding the soliton solutions of (\ref{cceqn}).

The $m$-CCNLS system exhibits a variety of interesting solitons due to the existence of additional coherent nonlinearities resulting from four-wave mixing process. Based on the presence and absence of coherent nonlinearities (respectively for $s\neq 0$ and $s=0$ in (\ref{3beq})), the obtained one-soliton solution (given in the Appendix \ref{one-sol-ccnls}) can be classified into two types, namely (i) coherently coupled solitons and (ii) incoherently coupled solitons, using the soliton parameters ($\alpha_u^{(j)}$, $u=1,2$, $j=1,2,3,...,m$). Particularly, in Ref. \cite{Kanna2011jpa}, it has been shown that for the choice $\sum_{j=1}^m (\alpha_u^{(j)})^2=0,~u=1,2$, the auxiliary function $s$ becomes zero and the corresponding one-soliton solution is said to be incoherently coupled soliton (ICS). But for the choice $\sum_{j=1}^m (\alpha_u^{(j)})^2\neq 0,~u=1,2$, $s$ becomes non-zero and the resulting one-soliton solution is said to be coherently coupled soliton (CCS). In general, ICS exhibits standard sech-type (single-hump) soliton profile  whereas the CCSs can have novel double-hump and flat-top profiles in addition to the single-hump (non-sech type) structures. A detailed analysis of these ICS and CCS is given in Ref.\cite{Kanna2011jpa}.

The main objective of the present paper is to give a clear picture about various energy sharing collisions of bright solitons in the above mentioned three integrable nonlinear systems, namely the Manakov, the mixed-ICNLS and the $m$-CCNLS systems. For this purpose, we make use of the soliton solutions obtained earlier and demonstrate the collisions graphically. We present the collision scenario of solitons in the Manakov and the mixed-CCNLS systems in Sec. \ref{sec-mana} and Sec. \ref{sec-mixed}, respectively. Bright soliton collision in the $m$-CCNLS system is given in Sec. \ref{sec-ccnls} and the final section is allotted for conclusion.

\section{Soliton collisions in the ICNLS (Manakov) system: Type-I energy sharing collision}\label{sec-mana}
To begin, we consider the collision of solitons in the celebrated Manakov system (Eqn. ({\ref{cnls}) with $\sigma=1$). Manakov himself has explicitly obtained one- and two- soliton solutions using the inverse scattering transform method. He has shown that in a two  soliton collision process, soliton polarization do not change only in the case when their initial polarizations are parallel or orthogonal. Later, Radhakrishnan {\it et al.} have shown that the solitons in the Manakov system (\ref{cnls}) exhibit certain novel inelastic (energy sharing) collisions \cite{RK1997pre} in contrast to single component NLS system. Kanna {\it et al}., have obtained the multisoliton solutions for the multicomponent Manakov system (\ref{mcnls}) using the Hirota's method \cite{Kanna2003pre}. Thus the system has been well studied in the literature \cite{Manakov,RK1997pre,Kannapramana,Kanna2001prl,Kanna2003pre,Kanna2006pre} and the existence of $N$-soliton (for arbitrary $N$) solution and also its proof has been obtained. In this section, we restrict our review to the interaction of two solitons in the Manakov system.

\subsection{Two-soliton solution and its collision dynamics}
The two-soliton solution of the Manakov system obtained by Radhakrishnan {\it et al.} \cite{RK1997pre} can be compactly written in terms of Gram determinant \cite{Kanna2009epjst} as
\bes \bea
q_j&&=\frac{g^{(j)}}{f}, \quad j=1, 2,
\label{trans}
\eea
where
\bea
\hspace{-0.5cm}g^{(j)}=
\left|
\begin{array}{ccccc}
A_{11} & A_{12}&1&0& e^{\eta_1}\\
A_{21} & A_{22}&0&1& e^{\eta_2}\\
-1&0 & B_{11} &B_{12} & 0\\
0&-1 & B_{21} &B_{22} & 0\\
0 &0& -\alpha_1^{(j)}&-\alpha_2^{(j)} & 0
\end{array}
\right|, \quad \quad f= \left|
\begin{array}{cccc}
A_{11} &A_{12}& 1&0\\
A_{21} &A_{22}& 0&1\\
-1&0 & B_{11}&B_{12} \\
0&-1 & B_{21}&B_{22} \\
\end{array}
\right|,
\label{Det2sol}
\eea \label{2sol-cnls}\ees
\noindent in which $A_{ij}=\ds{\frac{e^{\eta_i+\eta_j^*}}{k_i+k_j^*}}$, and $B_{ij}=\kappa_{ji}=\ds{\frac{\left(\alpha_j^{(1)}\alpha_i^{(1)*}+\alpha_j^{(2)}\alpha_i^{(2)*}\right)}{(k_j+k_i^*)}}$, \;\;\;$i,j=1,2$. The above most general bright two-soliton solution is characterized by six arbitrary complex parameters $k_1$, $k_2$, $\alpha_1^{(j)}$ and
$\alpha_2^{(j)}$, $j=1,2$ and it corresponds to the collision of two bright solitons.

Now, we discuss the collision dynamics of two bright solitons in the Manakov system.  One finds that enhancement and suppression of soliton intensities in different components occur as a consequence of energy exchange between the two colliding solitons as well as the two components.  This exchange phenomenon also satisfies the energy conservation of both the solitons before and after collision and also the conservation of energy in individual components which has been discussed in detail in Refs. \cite{RK1997pre,Kannapramana,Kanna2001prl,Kanna2003pre}. For illustrative purpose, we show the energy sharing collision characterized by intensity redistribution, amplitude dependent phase-shift and change in relative separation distances in the Manakov system in Fig. \ref{mana-sc1}. The parameters are chosen as $k_1=1+i, ~k_2=1.5-0.8i, \alpha_1^{(1)}=2.5,~ \alpha_1^{(2)}=0.2,~ \alpha_2^{(1)}=1.5$ and $\alpha_2^{(2)}=-0.6$. The two solitons $S_1$ and $S_2$ are well separated before and after collision in both the components $q_1$ and $q_2$. In the $q_1$ component  the intensity of soliton $S_1$ gets suppressed while that of soliton $S_2$ is enhanced after interaction and the reverse scenario takes place in the $q_2$ component.
\begin{figure}[h]
\centering
\includegraphics[width=0.37\linewidth]{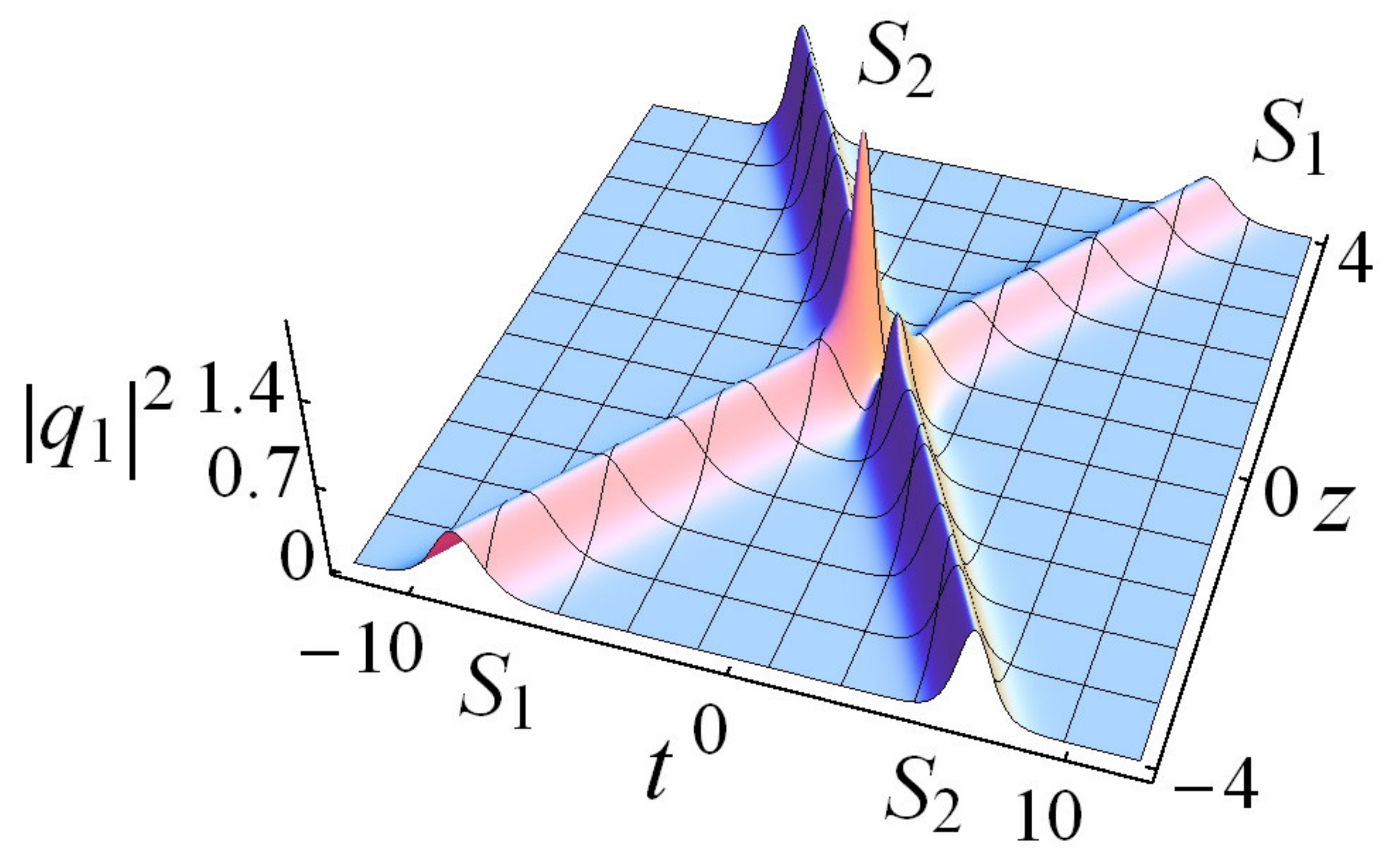}~~~~\includegraphics[width=0.37\linewidth]{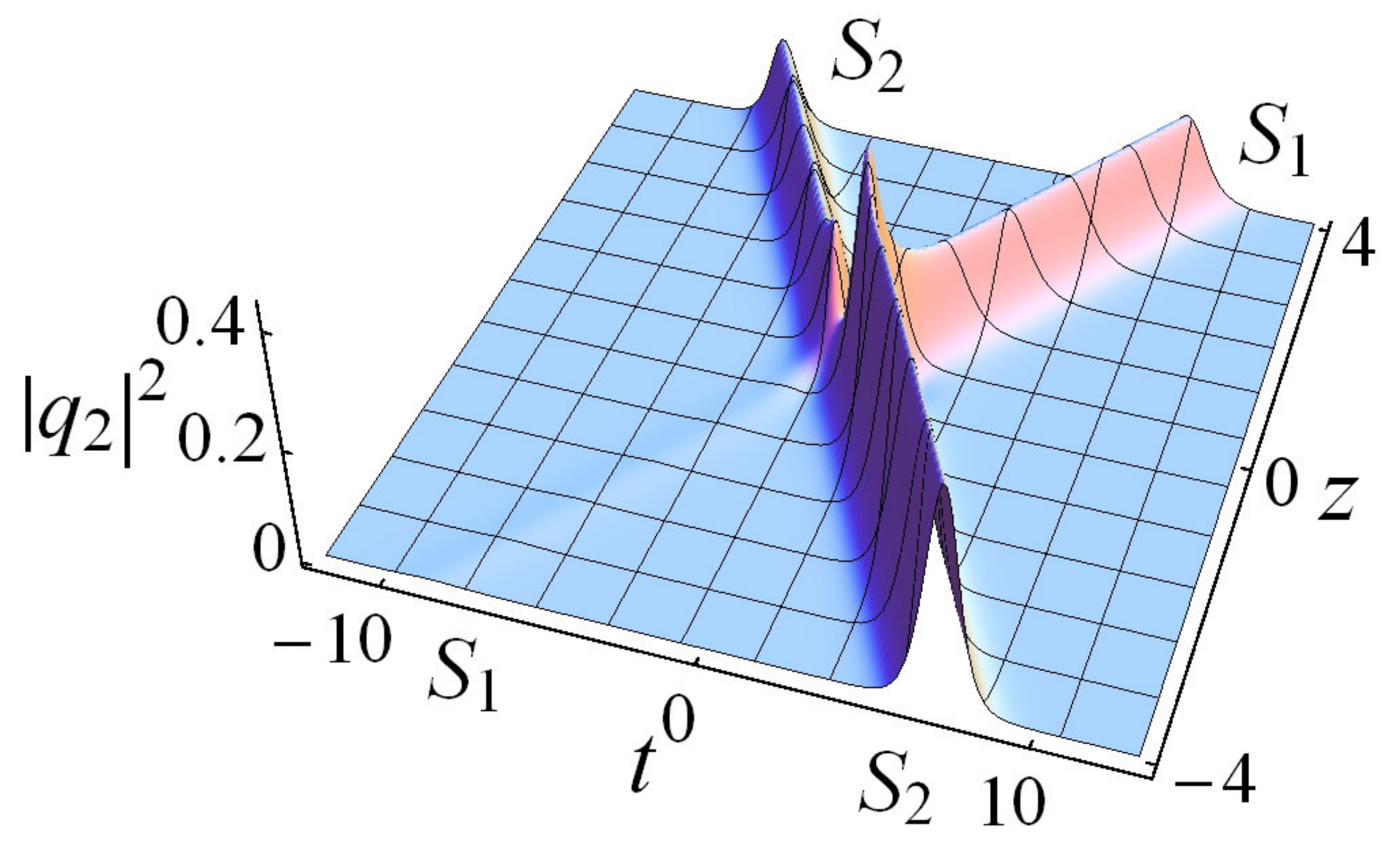}
\caption{Type-I energy sharing collision of Manakov solitons in $2$-ICNLS system.}
\label{mana-sc1}
\end{figure}

\subsection{Asymptotic analysis of two-soliton solution of Manakov system}
The understanding of this fascinating collision process can be facilitated
by making an asymptotic analysis of the two soliton solution of the Manakov case \cite{Kanna2003pre}. We perform the analysis
for the choice $k_{1R}, k_{2R}>0$ and $k_{1I}>k_{2I}$.  For any other choice
the analysis is similar. The study shows
that due to collision, the amplitudes of the colliding solitons $S_1$
and $S_2$
change from $(A_1^{1-}k_{1R},A_2^{1-}k_{1R})$ and
 $(A_1^{2-}k_{2R},A_2^{2-}k_{2R})$ to  $(A_1^{1+}k_{1R},
A_2^{1+}k_{1R})$ and  $(A_1^{2+}k_{2R},A_2^{2+}k_{2R})$, respectively.
Here  the superscripts in ${A_i^j}$'s denote the solitons (number(1,2)), the subscripts
represent the components (number(1,2)) and '$\pm$' signs stand for '$z \rightarrow \pm \infty$'.
One can find that
\bes \label{asy2sol} \bea A_j^{l+}=T_j^l A_j^{l-}, \quad j,l=1,2, \eea where
\bea
\left(
\begin{array}{c}
A_1^{1-}\\
A_2^{1-}
\end{array}
\right) & = &
\left(
\begin{array}{c}
\alpha_1^{(1)} \\
\alpha_1^{(2)}
\end{array}
\right)
\frac{e^{-R_1/2}}{(k_1+k_1^*)}, \\
\left(
\begin{array}{c}
A_1^{2-}\\
A_2^{2-}
\end{array}
\right)&=&
\left(
\begin{array}{c}
 e^{\delta_{11}}\\
e^{\delta_{12}}
\end{array}
\right) \frac{e^{-(R_1+R_3)/2}}{(k_2+k_2^*)},
\eea
and the transition amplitudes are given by
\bea
T_j^1&=&\left(\frac{(k_2+k_1^*)(k_1-k_2)}{(k_1+k_2^*)(k_1^*-k_2^*)}\right)^{\frac{1}{2}}
\left[\frac{1-\lambda_2}{\sqrt{1-\lambda_1\lambda_2}}\right], \;j=1,2,\\
T_j^2&=&-\left(\frac{(k_2+k_1^*)(k_1^*-k_2^*)}{(k_1-k_2)(k_1+k_2^*)}\right)^{\frac{1}{2}}
\left[\frac{\sqrt{1-\lambda_1\lambda_2}}{1-\lambda_1}\right], \;j=1,2.
\eea 
In the above expressions 
\bea
&&e^{R_j}=\frac{\kappa_{jj}}{k_j+k_j^*}, \quad 
e^{\delta_{1j}}=\frac{(k_1-k_2)(\alpha_1^{(j)}\kappa_{22}-\alpha_2^{(j)}\kappa_{21})}{(k_1+k_j^*)(k_2+k_j^*)}, \quad j=1,2,\\
&&e^{R_3}=\frac{|k_1-k_2|^2(\kappa_{11}\kappa_{22}-\kappa_{12}\kappa_{21})}{(k_1+k_1^*)(k_2+k_2^*)|k_1+k_2^*|^2},\\ 
&&\lambda_1=\frac{\kappa_{21}}{\kappa_{11}}\frac{\alpha_1^{(j)}}{\alpha_2^{(j)}}, \quad \lambda_2=\frac{\kappa_{12}}{\kappa_{22}}\frac{\alpha_2^{(j)}}{\alpha_1^{(j)}}, 
\eea
in which \bea &&\kappa_{ji}={\frac{\left(\alpha_j^{(1)}\alpha_i^{(1)*}+\alpha_j^{(2)}\alpha_i^{(2)*}\right)}{(k_j+k_i^*)}},\qquad  i,j=1,2.\eea\ees

In general, $|T_j^l|^2 \neq 1$ and hence there occurs intensity (energy) redistribution among the two colliding solitons as well as among the components. However, during the interaction process the total energy of each soliton is conserved, that is $|A_1^{l\pm}|^2+|A_2^{l\pm}|^2=1$, $l=1,2$. Another noticeable observation in this interaction process is that the intensity of each mode is separately conserved, that is $\int_{-\infty}^{\infty} |q_j|^2 dt =  \mbox{constant} , \;\;j=1,2$. Also, the colliding solitons $S_1$ and $S_2$ undergo amplitude dependent phase-shifts $\Phi_1$ and $\Phi_2$, respectively, given by
\bea
\Phi_1=-\Phi_2&=&\frac{1}{2}\ln\left[\frac{
|k_1-k_2|^2(\kappa_{11}\kappa_{22}-\kappa_{12}\kappa_{21})}
{|k_1+k_2^*|^2\kappa_{11}\kappa_{22}}\right].
\eea
Ultimately the above phase-shifts make the relative separation distance between the
solitons $t_{12}^{\pm}$ (position of $S_2$ (at $z \rightarrow \pm\infty$) minus position of
$S_1$  (at $z \rightarrow \pm \infty)$ also to vary during collision, depending
upon the amplitudes. The change in the relative separation distance is found to
be $\Delta t_{12}=t_{12}^--t_{12}^+ =\frac{(k_{1R}+k_{2R})}{k_{1R}k_{2R}} \Phi_1$.

We call such a collision scenario as type-I energy sharing collision (ESC). Such energy sharing collision occurs for $\frac{\alpha_1^{(1)}}{\alpha_2^{(1)}}\neq\frac{\alpha_1^{(2)}}{\alpha_2^{(2)}}$, which is quite general.  But when we choose $\frac{\alpha_1^{(1)}}{\alpha_2^{(1)}}=\frac{\alpha_1^{(2)}}{\alpha_2^{(2)}}$, the two solitons exhibit elastic collision only.   This interesting collision behaviour has also been experimentally verified in birefringent fibers \cite{ref30} and in photorefractive media \cite{ref16}.  The most important application of the energy sharing collision property is a theoretical possibility for constructing logic gates for optical computer.

\section{Soliton collisions in the mixed-ICNLS system: Type-II energy sharing collision}\label{sec-mixed}
Next, we consider the mixed 2-ICNLS system (\ref{cnls}) with $\sigma=-1$. This system (\ref{cnls}) admits three types of soliton solutions namely bright-bright, bright-dark, and dark-dark. It was found that the bright solitons exhibit a special type of energy sharing collisions where as the dark solitons always undergo elastic collision \cite{Kanna2006pre,Kanna2008pra}. The Gram determinant form of the bright $N$-soliton solution (for arbitrary $N$) has been obtained by Kanna et al., \cite{Kanna2006pre}. In this section, we revisit the collision dynamics of two bright solitons in detail. For this purpose, we consider the bright two-soliton solution of the mixed 2-CNLS equations which is given by Eq. (\ref{2sol-cnls}) with the redefinition of $B_{ij}$ as
\bea
B_{ij}=\kappa_{ji}= \frac{(\alpha_j^{(1)}\alpha_i^{(1)*}-
\alpha_j^{(2)}\alpha_i^{(2)*})}{\left(k_j+k_i^*\right)},\;i,l=1,2.\label{mixed_kappa}
\end{eqnarray}
Note that the form of the above two-soliton solution remains the same as that of the Manakov case except for the crucial difference in the expressions for $\kappa_{ji}$.

It was found that mixed 2-CNLS equations admit energy sharing collision of bright solitons in a quite different manner from the collision scenario of the Manakov system \cite{Kanna2006pre}. It has been shown that in mixed CNLS equations during a two soliton collision process  there is a possibility of either enhancement or suppression of intensity in a given soliton in all the components \cite{Kanna2006pre}. Here also the collision process is characterized as in the focusing case.   The most important consequence of the above energy sharing collision is the possibility of {\it{soliton amplification}} in all the components.  Fig. \ref{2c-mixed} shows that after collision  the first soliton $S_1$ in the component $q_1$ gets enhanced in its amplitude while the soliton $S_2$ is suppressed.  Interestingly, the same kind of changes are observed in the second component $q_2$ as well.  As the two-soliton solution of mixed ICNLS equation is same as that of the Manakov system except for $\kappa_{ji}$, the asymptotic expressions are also same as given by (\ref{asy2sol}) with $\kappa_{il}$ as given in Eq. (\ref{mixed_kappa}). The analysis reveals the fact that the colliding solitons change their amplitudes in each component according to the conservation equation
\bea
 |A_1^{j-}|^2 - |A_2^{j-}|^2 =  |A_1^{j+}|^2 - |A_2^{j+}|^2 =1,\;\; j=1,2, \label{12}
 \eea
where $A_j^{l\pm}$ are given by Eqn. (\ref{asy2sol}) with modified $\kappa_{ji}$s. This condition allows the given soliton to experience the same effect
in each component during collision, which may find potential application
in the noiseless amplification of a pulse.
It can be easily observed from the conservation relation (\ref{12})
that each component of a given soliton
experiences the same kind of energy switching during collision process.
The other soliton (say $ S_2$ ) experiences an opposite kind of energy
switching due to the conservation law $\int_{-\infty}^{\infty} |q_j|^2 dt =  \mbox{constant} , \;\;j=1,2$.
\begin{figure}[h]
\centering\includegraphics[width=0.38\linewidth]{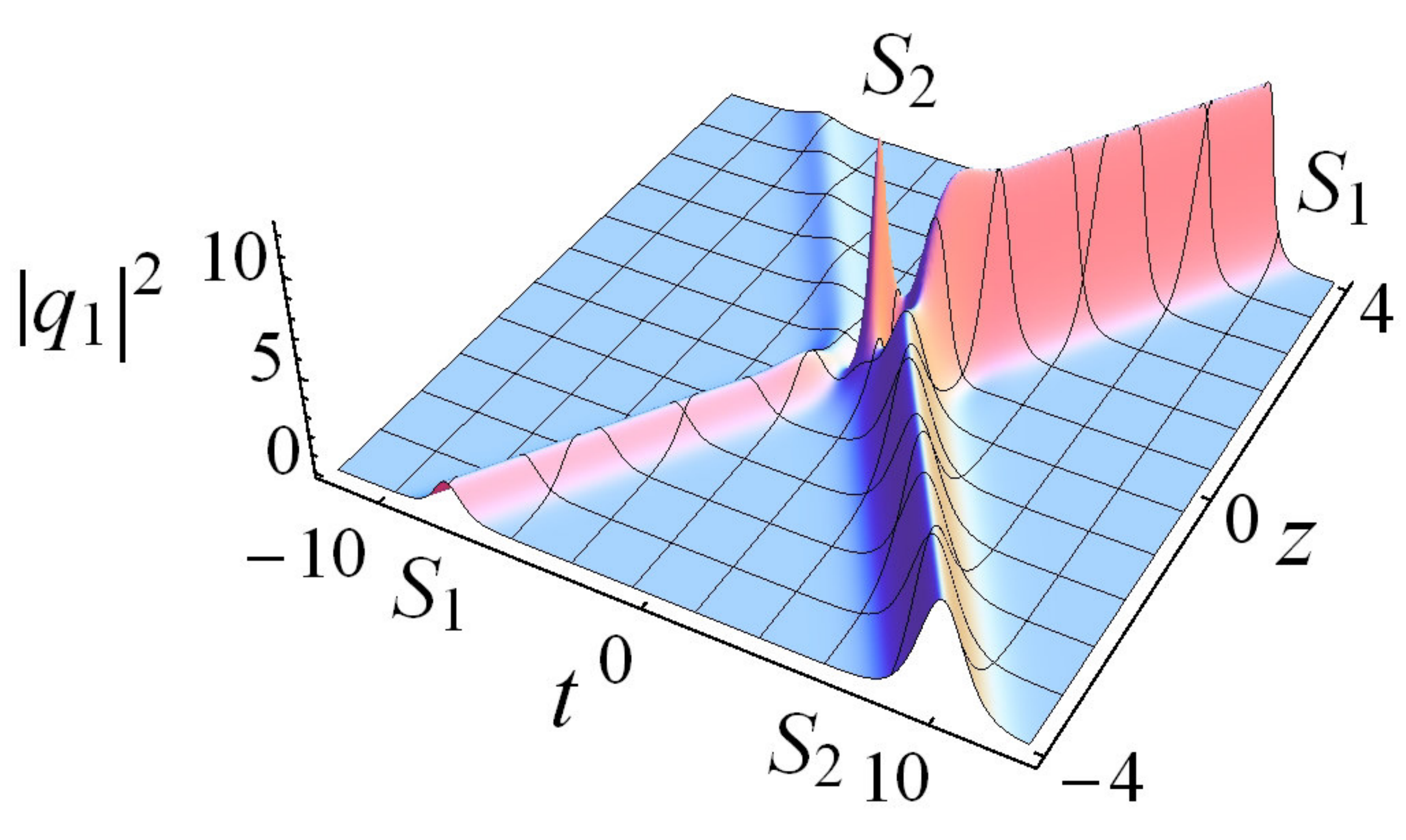}~~~~\includegraphics[width=0.37\linewidth]{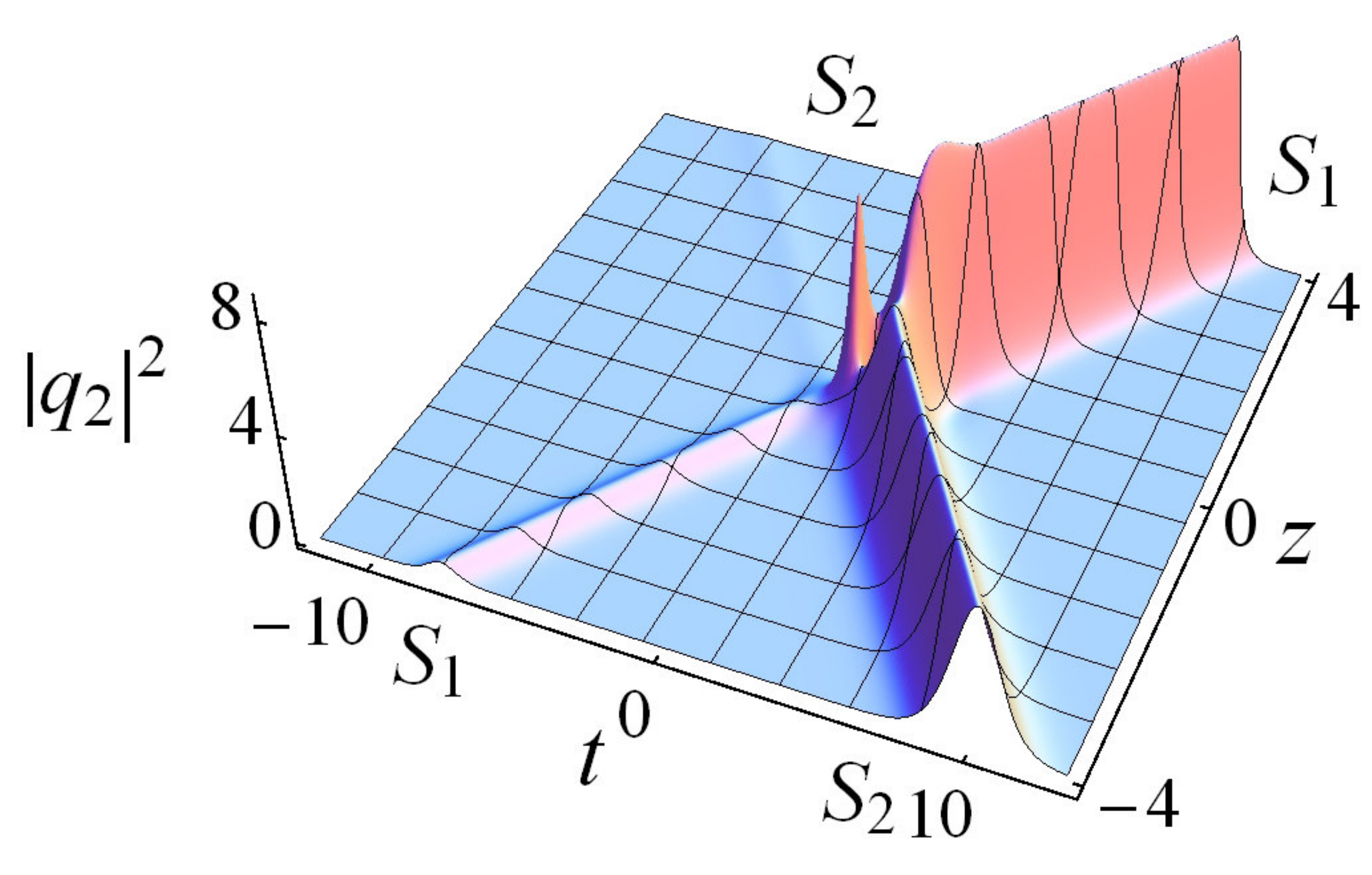}
\caption{Type-II energy sharing collision of bright solitons in mixed $2$-ICNLS system for $k_1=1.5+i, ~k_2=1-i, \alpha_1^{(1)}=1+i,~ \alpha_1^{(2)}=0.8+0.2i,~ \alpha_2^{(1)}=1-i$ and $\alpha_2^{(2)}=0.5$.}
\label{2c-mixed}
\end{figure}

For the standard elastic collision  property ascribed to the scalar solitons
to occur here we need the magnitudes of the transition intensities to be unity which is possible for
the specific choice $\frac{\alpha_1^{(1)}}{\alpha_2^{(1)}}=\frac{\alpha_1^{(2)}}{\alpha_2^{(2)}}$.
The other quantities characterizing this collision process, along with
this energy redistribution, are the amplitude dependent phase-shifts
and change in relative separation distances. The corresponding  expressions take the same form as that of the Manakov model with the redefinition of $\kappa_{ji}$ as in eq. (\ref{12}). We refer to this collision process in which a given soliton experiences same kind of energy switching in both components as type-II energy sharing collision.

Thus the type-II energy sharing collision scenario is
entirely different from the one observed in the Manakov system  where
one soliton gets suppressed in one component and is enhanced in the other
component with commensurate changes in the other soliton. The collision scenario shown in Fig. \ref{2c-mixed} can also be viewed as an amplification process in which the soliton $S_1$ represents a signal or data carrier while the soliton $S_2$ represents an energy reservoir or pump.  The main advantage of this amplification process is that it does not require any external amplification medium and therefore the amplification of $S_1$ does not introduce any noise \cite{Kanna2006pre}.

\section{Soliton collisions in the $m$-CCNLS equations: Type-III and Type-IV energy sharing collisions}\label{sec-ccnls}
The bright solitons of the $m$-CCNLS system (\ref{cceqn}) display interesting collision properties. Unlike in the other multicomponent nonlinear systems, namely Manakov system and mixed-ICNLS system, the $m$-CCNLS system shows novel energy switching mechanism for a given soliton while the other soliton reappears elastically after collision. The exact two soliton solution describing different types of collision scenario in $m$-CCNLS system is given in Appendix B. Importantly, based on the one-soliton solution, we can classify the two soliton collisions of the $m$-CCNLS system into three cases, namely
\newline (i) collision between a CCS and an ICS $\left(\sum_{j=1}^m (\alpha_1^{(j)})^2 \neq 0\right.$ and $\left.\sum_{j=1}^m (\alpha_2^{(j)})^2 =0\right)$,
\newline (ii) collision between two CCSs $\left(\sum_{j=1}^m (\alpha_u^{(j)})^2 \neq 0,~u=1,2\right)$, and
\newline (iii) collision between two ICSs $\left(\sum_{j=1}^m (\alpha_u^{(j)})^2=0,~u=1,2\right)$.\\
In order to understand these collision dynamics more clear, we have performed an asymptotic analysis of the two-soliton solution given in the Appendix with $k_{1R},~k_{2R}>0$ and $k_{1I}>k_{2I}$. In the following, we discuss various combinations of soliton collisions of the $m$-CCNS system (\ref{cceqn}) for $m\geq 2$. Here and in the following, the two colliding solitons are represented as $S_1$ and $S_2$.

\subsection*{(i) Collision between a CCS and ICS: Type-III energy sharing collision}
Let us consider the collision of two bright solitons, in which $S_1$ is of CCS type while $S_2$ is an ICS. From the detailed asymptotic analysis, which we skipped here, the amplitude of the given CCS $S_1$ and ICS $S_2$ before collision ($A_j^{u-}$) can be related to that of after collision ($A_j^{u+}$), by the transition amplitudes ($T_j^{(u)}$) as,
\bes\bea
A_j^{u+}&=&T_j^{(u)}~A_j^{u-}, \qquad  u=1,2, \quad j=1,2,3,...,m, \qquad\qquad\qquad\qquad \qquad\\
\hspace{-1.5cm}\mbox{where}  \qquad\qquad &&\qquad\qquad\nonumber\\
\hspace{-1.5cm}T_j^{(1)}&=&\left(\frac{(k_1^*+k_2)(k_1-k_2)\big|(\alpha_1^{(j)} \kappa_{22}-\alpha_2^{(j)} \kappa_{12})+\alpha_2^{(j)*} \Omega\big|^2}{(k_1+k_2^*) (k_1^*-k_2^*) ~\kappa_{22}^2 ~ |\alpha_1^{(j)}|^2}\right)^{\frac{1}{2}}, \\
\hspace{-1.5cm}T_j^{(2)}&=&\frac{(k_1^* + k_2)(k_1^* - k_2^*)}{(k_1 - k_2)(k_1 + k_2^*)}, ~~ j=1,2,3,...,m,
\eea\label{tr3c}\ees
in which $\Omega=\frac{\gamma \sum_{j=1}^m (\alpha_1^{(j)}\alpha_2^{(j)})}{(k_1-k_2)}$ and $\kappa_{uv}=\frac{\gamma}{(k_u+k_v^*)}\sum_{j=1}^m(\alpha_u^{(j)} \alpha_v^{(j)*})$, $u,v=1,2$. The above mentioned transition amplitudes ($T_j^{(u)}$) determine the collision nature of a given soliton $S_u$, $u=1,2$, in a particular component $q_j$, $j=1,2,3,...,m$.

From the above equation (\ref{tr3c}), one can understand that the solitons undergo elastic collision when their transition intensities become uni-modular, which results in the solitons with same intensities before and after collision. Especially, the ICS $S_2$ exhibits elastic collision always as $|T_j^{(2)}|^2=1$ without any restriction on the soliton parameters. However, the CCS $S_1$ undergoes energy switching collision for general choice of soliton parameters. Only for specific choice $\mbox{when}~\sum_{j=1}^m (\alpha_1^{(j)})^2 = 0$, one can expect elastic collision, but this is not possible as this choice restricts $S_1$ to be an ICS. Here we observe that in a given component one soliton retains its intensity while the other undergoes change in its intensity after collision, which shows the non-conservation of energy in that component. In order to conserve the total energy of the system, the corresponding soliton undergoes opposite kind of intensity switching in another component. Apart from the change or invariance in the intensity/amplitude, the colliding solitons CCS $S_1$ and ICS $S_2$ experience phase-shifts $\Phi_1= \frac{1}{k_{1R}}\ln\left(\frac{(k_1-k_2)(k_1^*-k_2^*)}{(k_1+k_2^*)(k_1^*+k_2)}\right)$ and $\Phi_2= -\left(\frac{2k_{1R}}{k_{2R}}\right)\Phi_1$, respectively after collision. This will result in a change in the relative separation distance between the solitons before collision ($t_{12}^- =\frac{\theta_{11}-\epsilon_{11}}{2k_{2R}}-\frac{\epsilon_{11}}{4k_{1R}}$: position of soliton $S_2$ minus position of soliton $S_1$ before collision) and after collision ($t_{12}^+ =\frac{R_2}{2k_{2R}}-\frac{\theta_{11}-R_2}{4k_{1R}}$: position of soliton $S_2$ minus position of soliton $S_1$ after collision) and this can be written as $\Delta t_{12}=t_{12}^- - t_{12}^+=\left(1+\frac{2 k_{1R}}{k_{2R}}\right) \Phi_1$. Except the transition amplitudes $T_j^{(u)}$, both the phase-shift and the relative separation distance are independent of $\alpha_u^{(j)}$ parameters. Note that the reverse type of energy switching scenario is also possible for CCS $S_2$ in the two components which can be obtained for proper choice of $\alpha_u^{(j)}$. We refer to this energy sharing collision with energy switching occurring in CCS only with opposite nature in the two components $q_1$ and $q_2$ as type-III energy sharing collision.

To be more clear, we explicitly demonstrate the CCS-ICS collision in 2-CCNLS and 3-CCNLS systems, which can be generalized to $m$-CCNLS system, with $m> 3$. In Fig. \ref{2c-ccsics}, we have shown the energy switching collision of CCS $S_1$ with ICS $S_2$ for the choice $k_1=2.3+i, ~k_2=2.5-i, ~\gamma=2,~ \alpha_1^{(1)}=0.75i,~ \alpha_1^{(2)}=1.9,~ \alpha_2^{(1)}=1+i$ and $\alpha_2^{(2)}=1-i$. Here CCS $S_1$ changes its profile from a double-hump (single-hump) to a single-hump (double-hump) structure with enhancement (suppression) of intensity in the $q_1$ ($q_2$) component, but the ICS $S_2$ exhibits elastic collision in both components.

\begin{figure}[h]
\centering\includegraphics[width=0.37\linewidth]{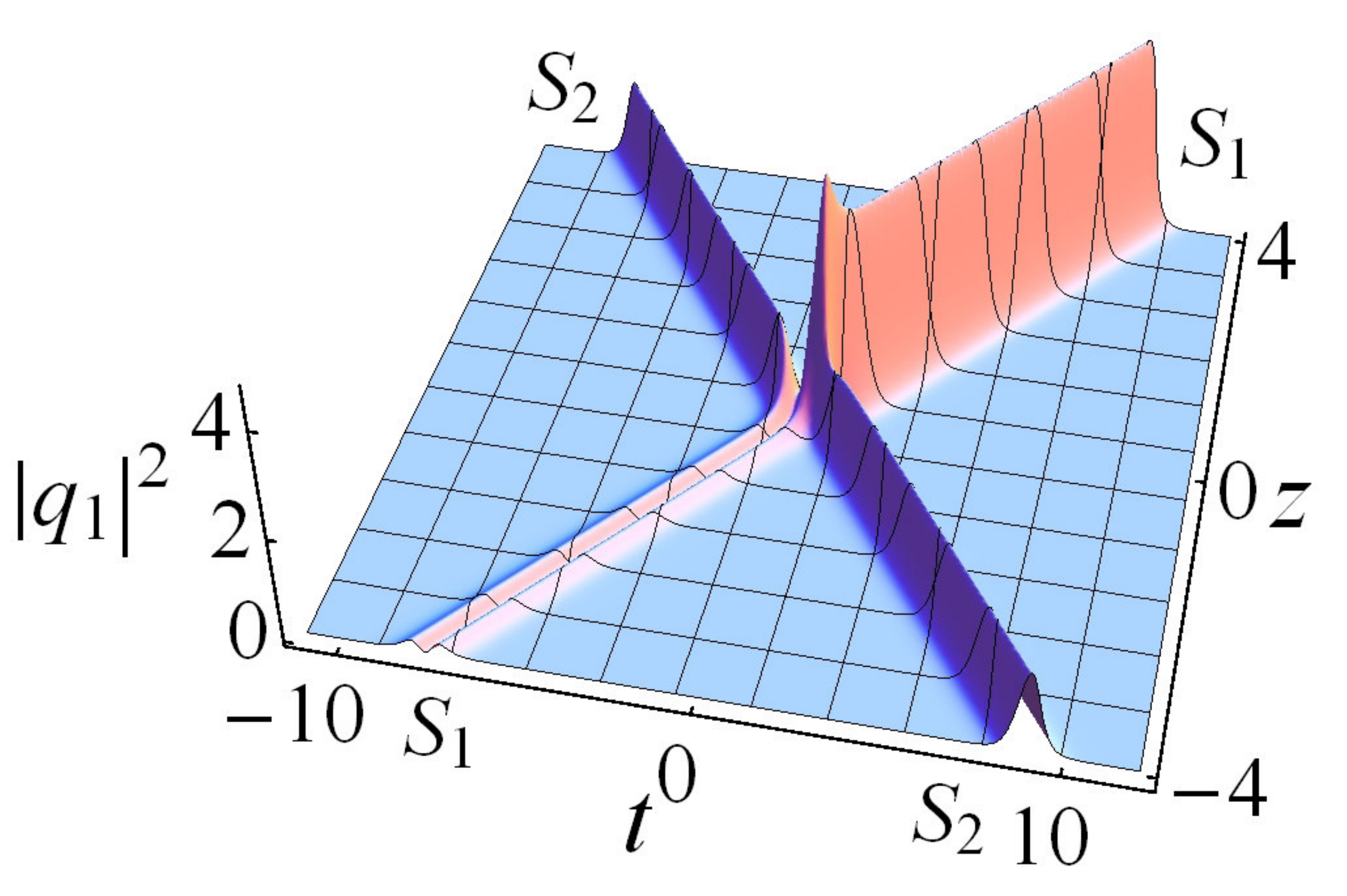}~~~~\includegraphics[width=0.37\linewidth]{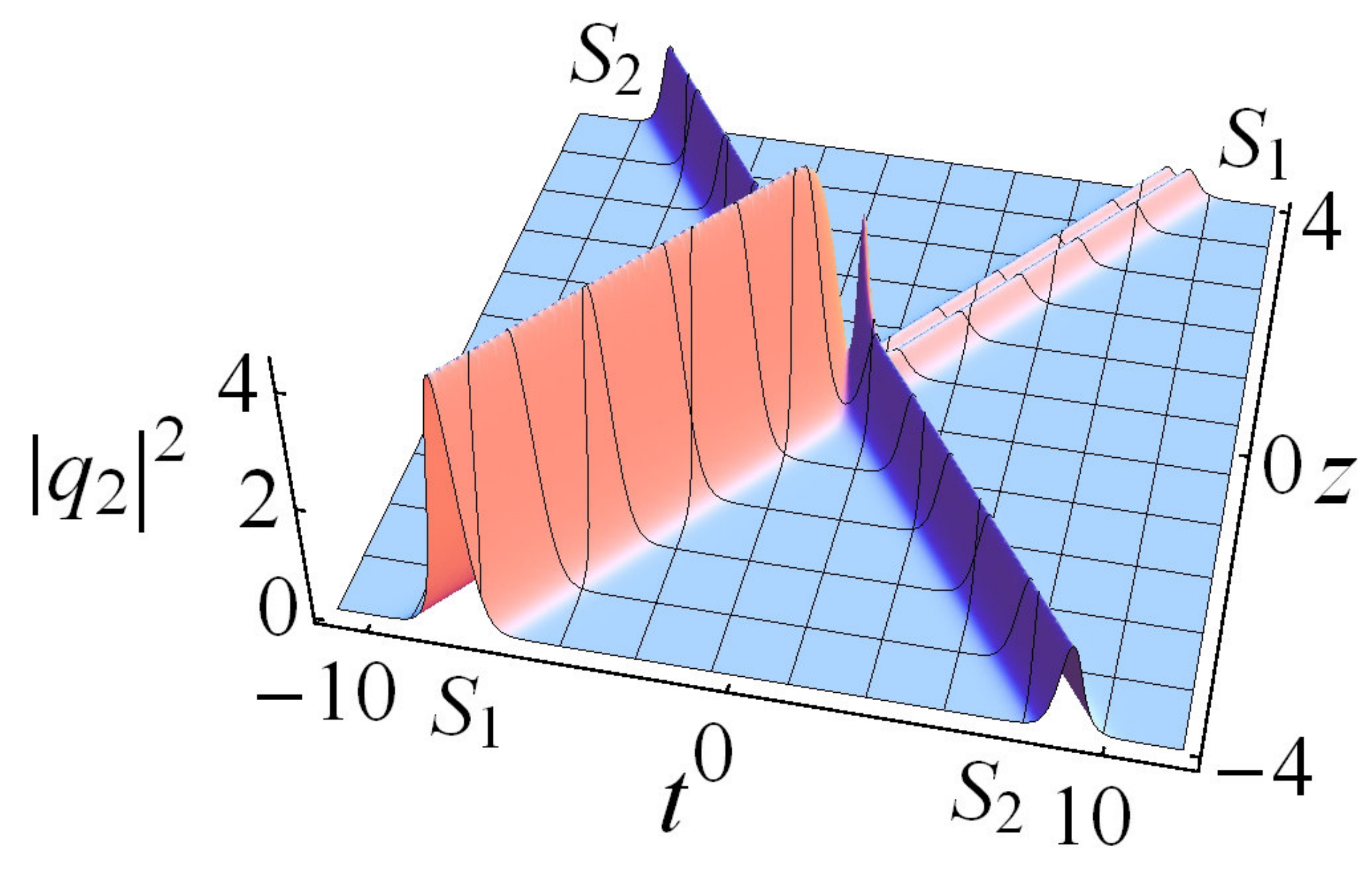}
\caption{Type-III energy sharing collision of CCS with ICS in $2$-CCNLS system.}
\label{2c-ccsics}
\end{figure}
It should be noted that the present CCNLS system (\ref{cceqn}) conserves total energy but the energy in the individual component is not conserved. In order to understand this, we obtain the form of energy conservation from (\ref{2cceqn}), simply for 2-CCNLS case as $i\frac{d}{dz}\int_{-\infty}^{\infty}|q_1|^2~dt = \gamma \int_{-\infty}^{\infty} (q_1^{*2}q_2^2-q_1^2q_2^{*2})~dt$ and $i\frac{d}{dz}\int_{-\infty}^{\infty}|q_2|^2~dt = \gamma \int_{-\infty}^{\infty} (q_1^2q_2^{*2}-q_1^{*2}q_2^2)~dt$. This shows that the energy in individual component is not conserved ($\frac{d}{dz}\int_{-\infty}^{\infty}|q_j|^2~dt \neq 0$, $j=1,2$) but the total energy is conserved ($\frac{d}{dz}\int_{-\infty}^{\infty}(|q_1|^2+|q_2|^2)~dt=0$). As a consequence of this, the ICS induces significant energy switching in the CCS with an amplitude dependent phase shift and reappears elastically after interaction.

The 3-CCNLS system admits more possible ways of energy switching for CCS. One possible way is depicted in Fig. \ref{3c-ccsics} for  $k_1=1.5+i, ~k_2=2-i, ~\gamma=2,~ \alpha_1^{(1)}=1,~ \alpha_1^{(2)}=1.5,~\alpha_1^{(3)}=2,~ \alpha_2^{(1)}=2+i$, $\alpha_2^{(2)}=2-i$, and $\alpha_2^{(3)}=\sqrt{6}~i$. It is evident from Fig. \ref{3c-ccsics}, the CCS $S_1$ changes its profile from single-hump to double-hump with suppression in its intensity in the $q_1$ and $q_3$ components. However, in the $q_2$ component, CCS $S_1$ just increases its intensity without change in the nature of profile. Similar to 2-CCNLS system, in 3-CCNLS system too the ICS $S_2$ remains same before and after collision in all the three components. In fact, one can have various combinations of energy switching collision of CCS with ICS for different choices of soliton parameters.
\begin{figure}[h]
\centering\includegraphics[width=0.33\linewidth]{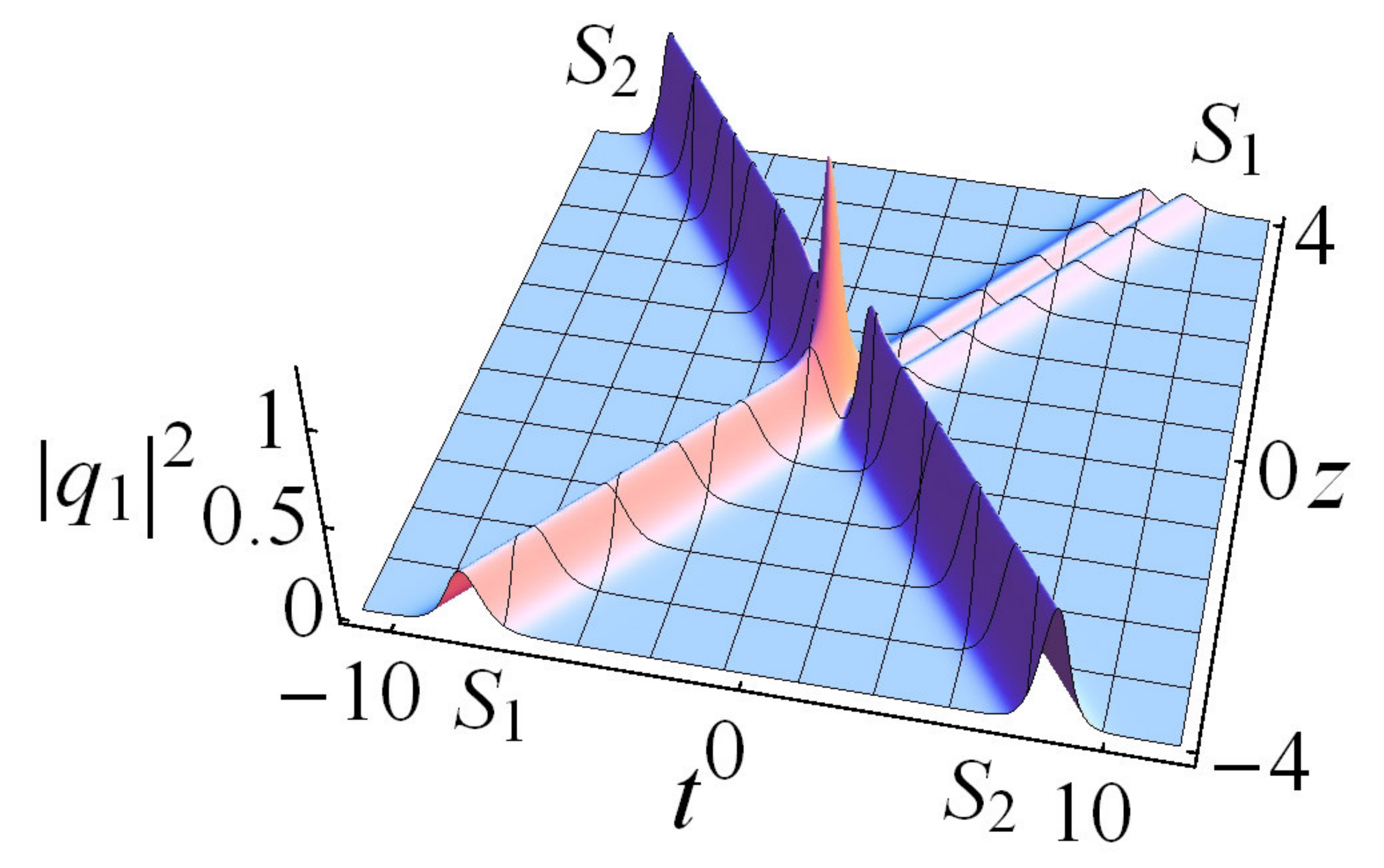}~~\includegraphics[width=0.33\linewidth]{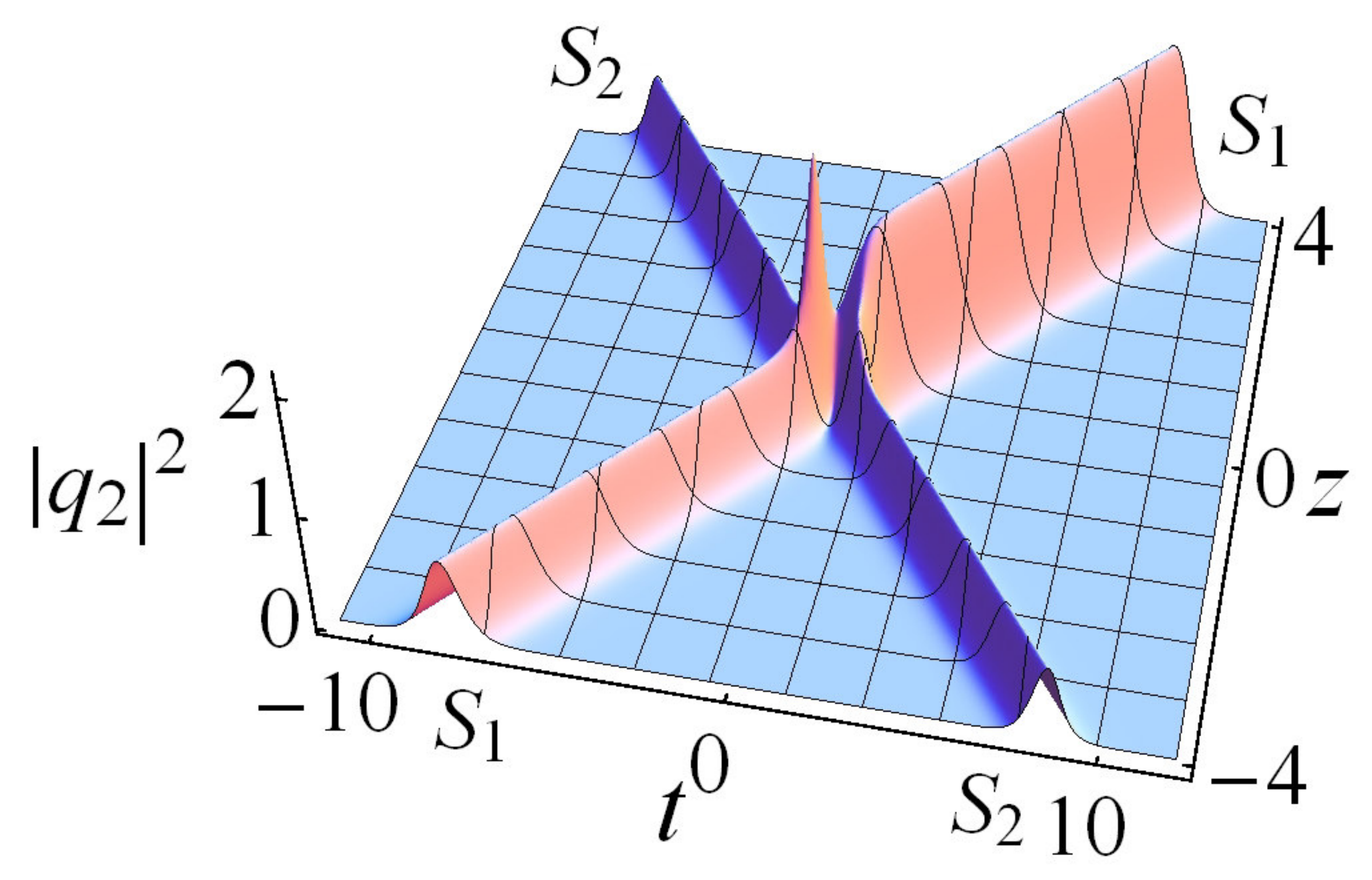}~~~~\includegraphics[width=0.33\linewidth]{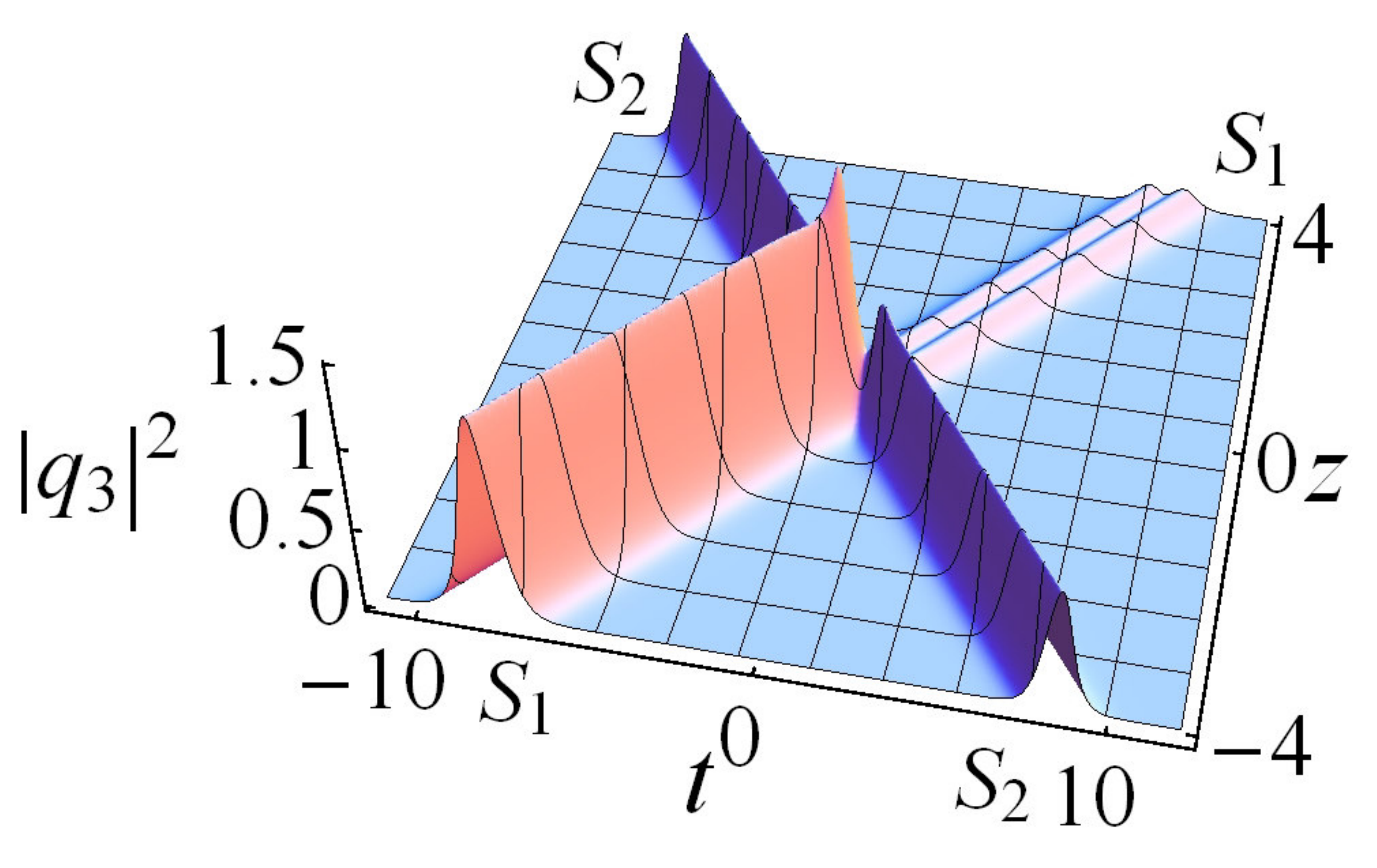}
\caption{Type-III energy sharing collision of CCS with ICS in $3$-CCNLS system.}
\label{3c-ccsics}
\end{figure}

\subsection*{(ii) Collision of two CCSs:}
Here the two CCSs can be obtained for the choice $\sum_{j=1}^m (\alpha_u^{(j)})^2 \neq 0,~u=1,2$. Then the relations between the amplitudes of solitons before and after collision are obtained by an asymptotic analysis of two-soliton solution (\ref{two-sol-ccnls-a}) as $A_j^{1+}= \frac{(k_1-k_2)(k_1^*+k_2)}{(k_1^*-k_2^*)(k_1+k_2^*)}A_j^{1-}$ and $A_j^{2+}= \frac{(k_1^*-k_2^*)(k_1^*+k_2)}{(k_1-k_2)(k_1+k_2^*)}A_j^{2-}$, $j=1,2,3,...,m$. From these expressions, we can easily find that the CCSs always undergo elastic collision with different profile structures as the corresponding relations for intensities become $|A_j^{u+}|^2=|A_j^{u-}|^2,~u=1,2,~j=1,2,3,...,m$. But these solitons, CCS $S_1$ and CCS $S_2$, exhibit phase-shifts after collision $\Phi_1= \frac{1}{k_{1R}}\ln\left(\frac{(k_1-k_2)(k_1^*-k_2^*)}{(k_1+k_2^*)(k_1^*+k_2)}\right)$ and $\Phi_2=-\left(\frac{k_{1R}}{k_{2R}}\right)\Phi_1$, respectively with a change in the relative separation distance ($\Delta t_{12}= \left(1+\frac{k_{1R}}{k_{2R}}\right) \Phi_1$) between the two CCSs.

For illustrative purpose, we have shown the collision between two CCSs in the 2-CCNLS and 3-CCNLS systems, respectively in Fig. \ref{2c-ccsccs} and Fig. \ref{3c-ccsccs} for the choice $k_1=1.5+i, ~k_2=2-i, ~\gamma=2,~ \alpha_1^{(1)}=1.7i,~ \alpha_1^{(2)}=1,~ \alpha_2^{(1)}=2i$, $\alpha_2^{(2)}=1.2$ and $k_1=1.5+i, ~k_2=2-i, ~\gamma=2,~ \alpha_1^{(1)}=0.25,~ \alpha_1^{(2)}=-0.71,~\alpha_1^{(3)}=1.2i,~ \alpha_2^{(1)}=1$, $\alpha_2^{(2)}=1.4i$, $\alpha_2^{(3)}=0.75i$. In Fig. \ref{2c-ccsccs}, two CCSs having single-hump (double-hump) profiles in $q_1$ ($q_2$) component undergo elastic collision. In Fig. \ref{3c-ccsccs}, the collision takes place between two double-hump CCSs in $q_1$, a single-hump and double-hump CCSs in $q_2$ and two single-hump CCSs in $q_3$ components. Other combinations of soliton profiles for the elastic collision of CCSs can also be achieved by tuning the $\alpha_u^{(j)}$ parameters.

\begin{figure}[h]
\centering\includegraphics[width=0.33\linewidth]{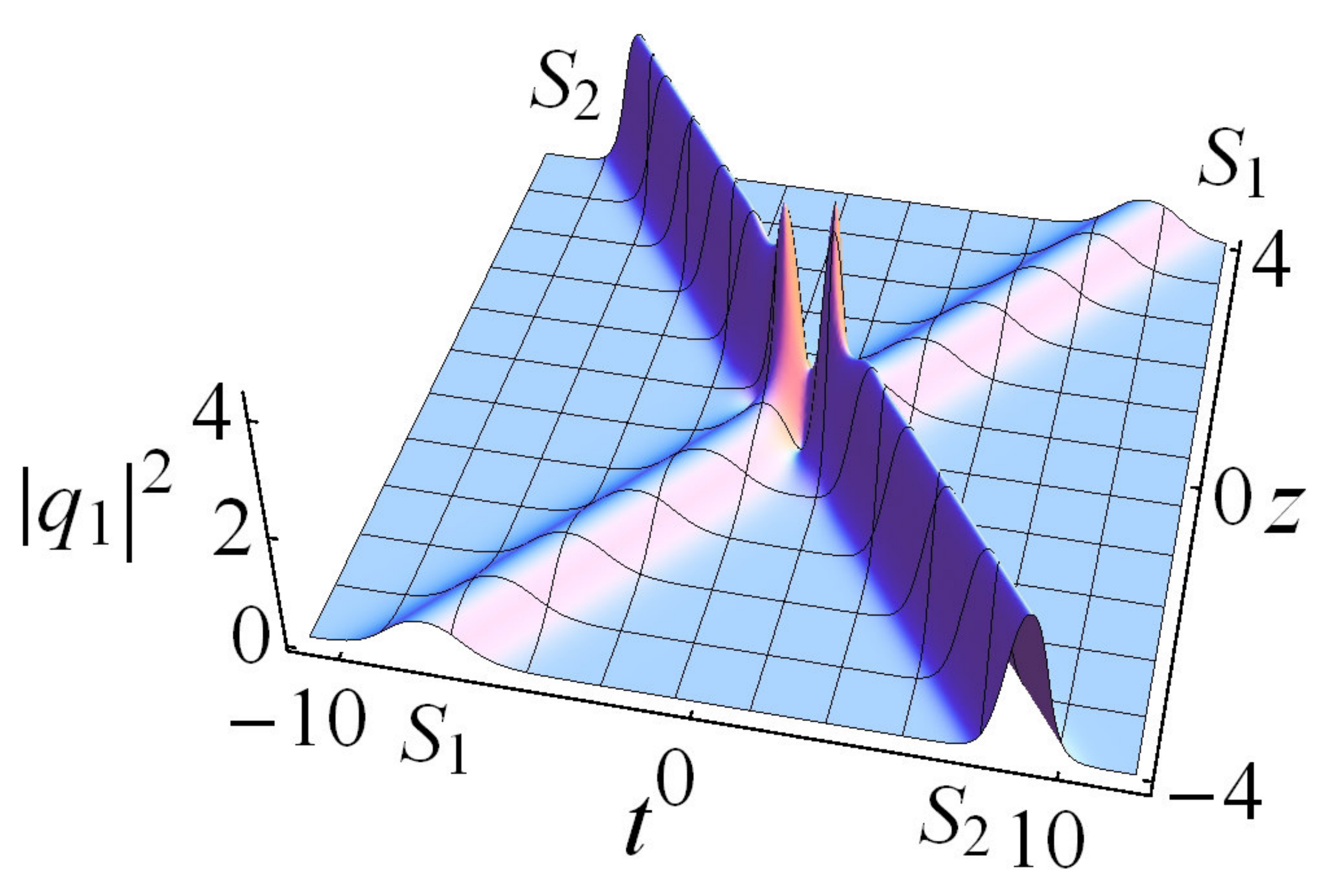}~~~~\includegraphics[width=0.33\linewidth]{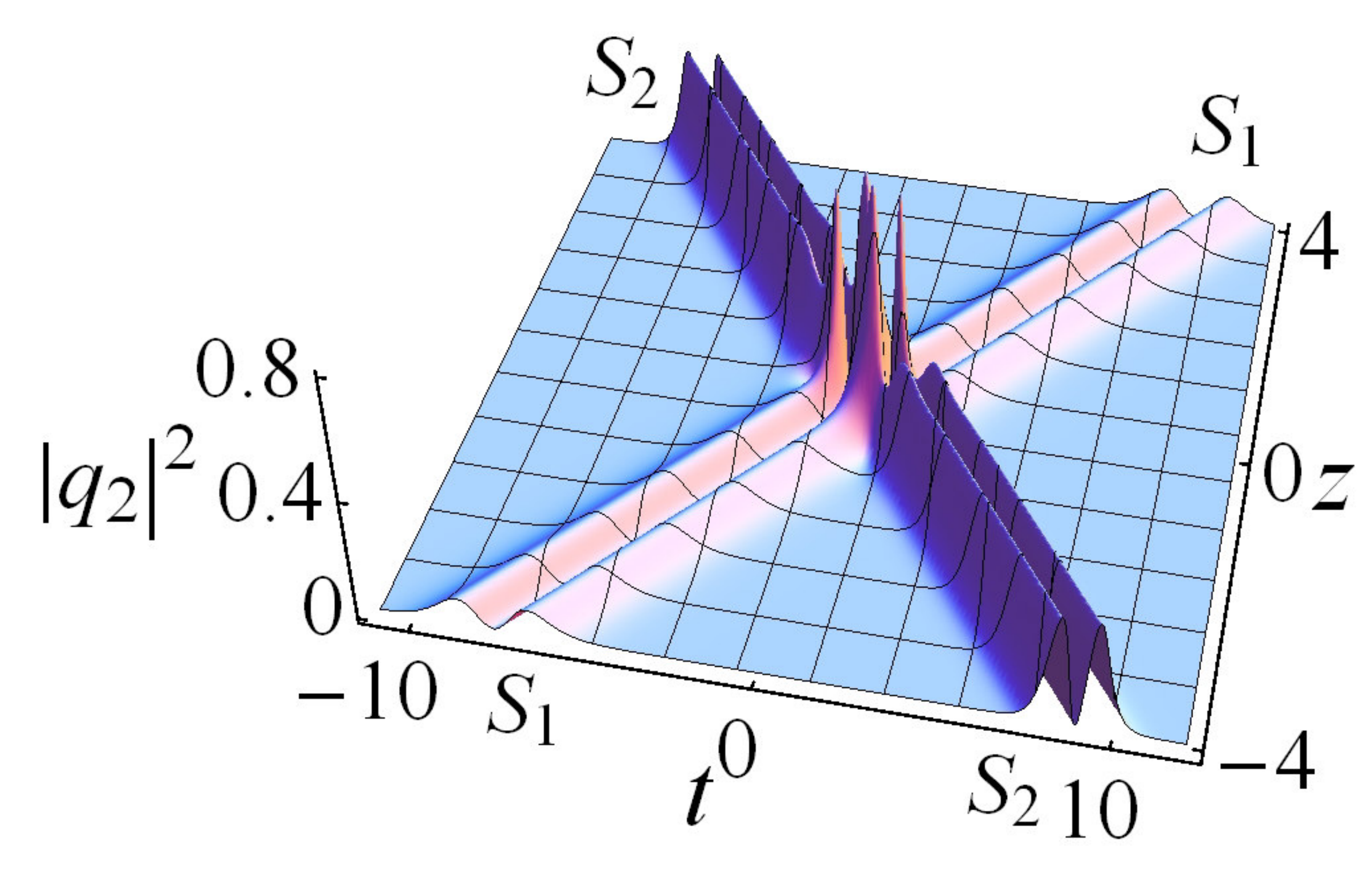}
\caption{Elastic collision of coherently coupled solitons in $2$-CCNLS system.}
\label{2c-ccsccs}
\centering
\includegraphics[width=0.33\linewidth]{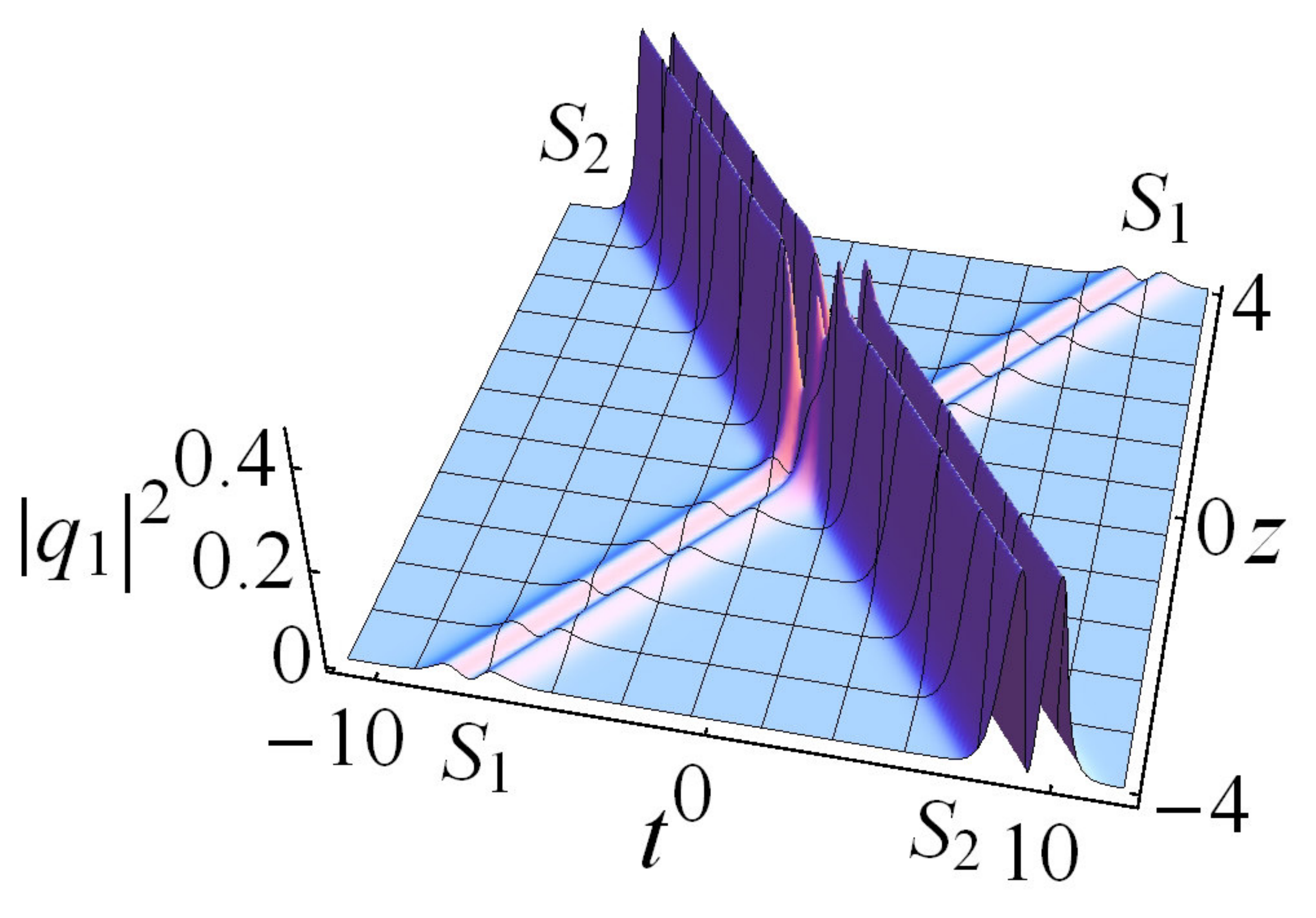}~~~~\includegraphics[width=0.34\linewidth]{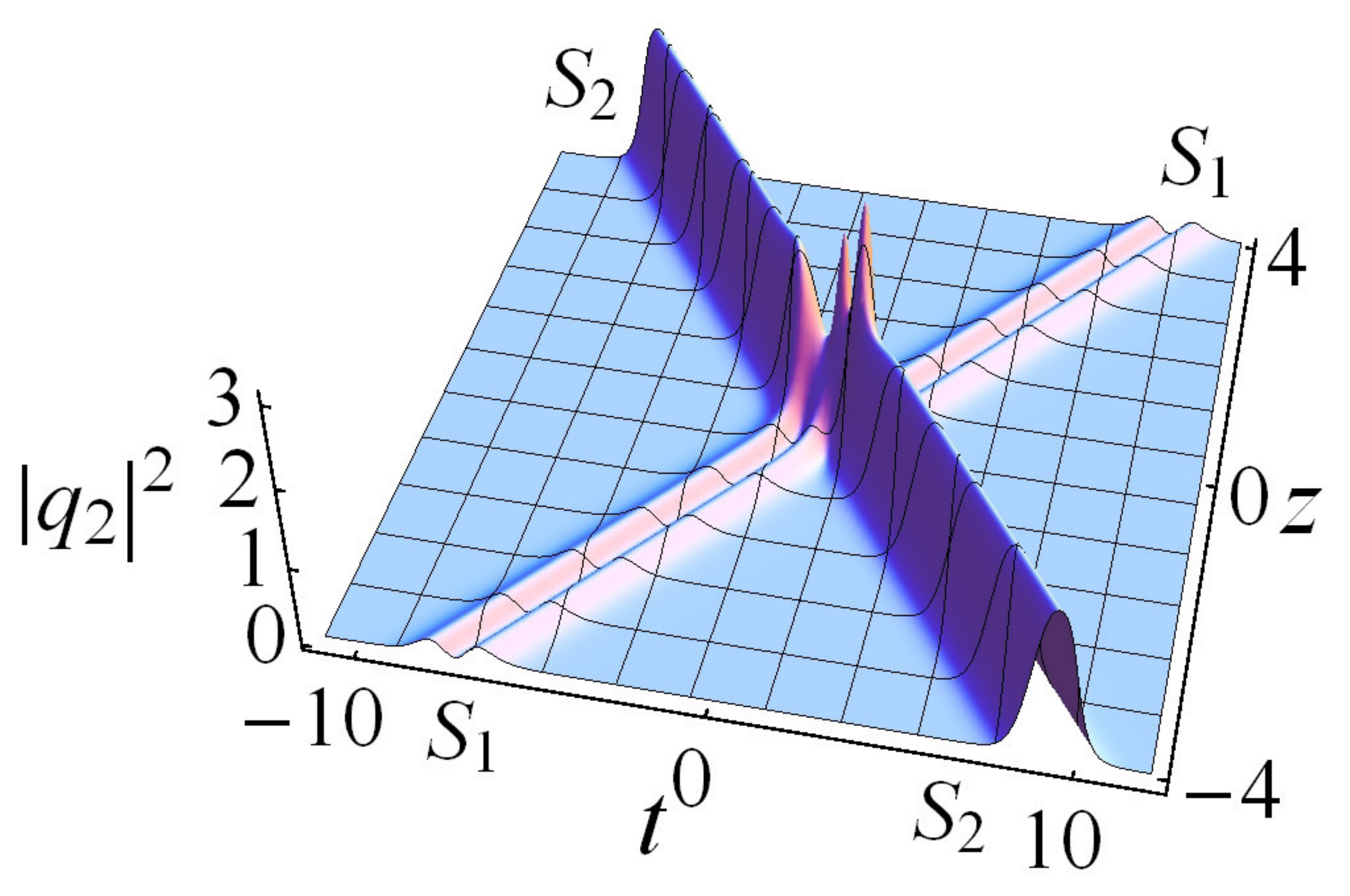}~~~~\includegraphics[width=0.34\linewidth]{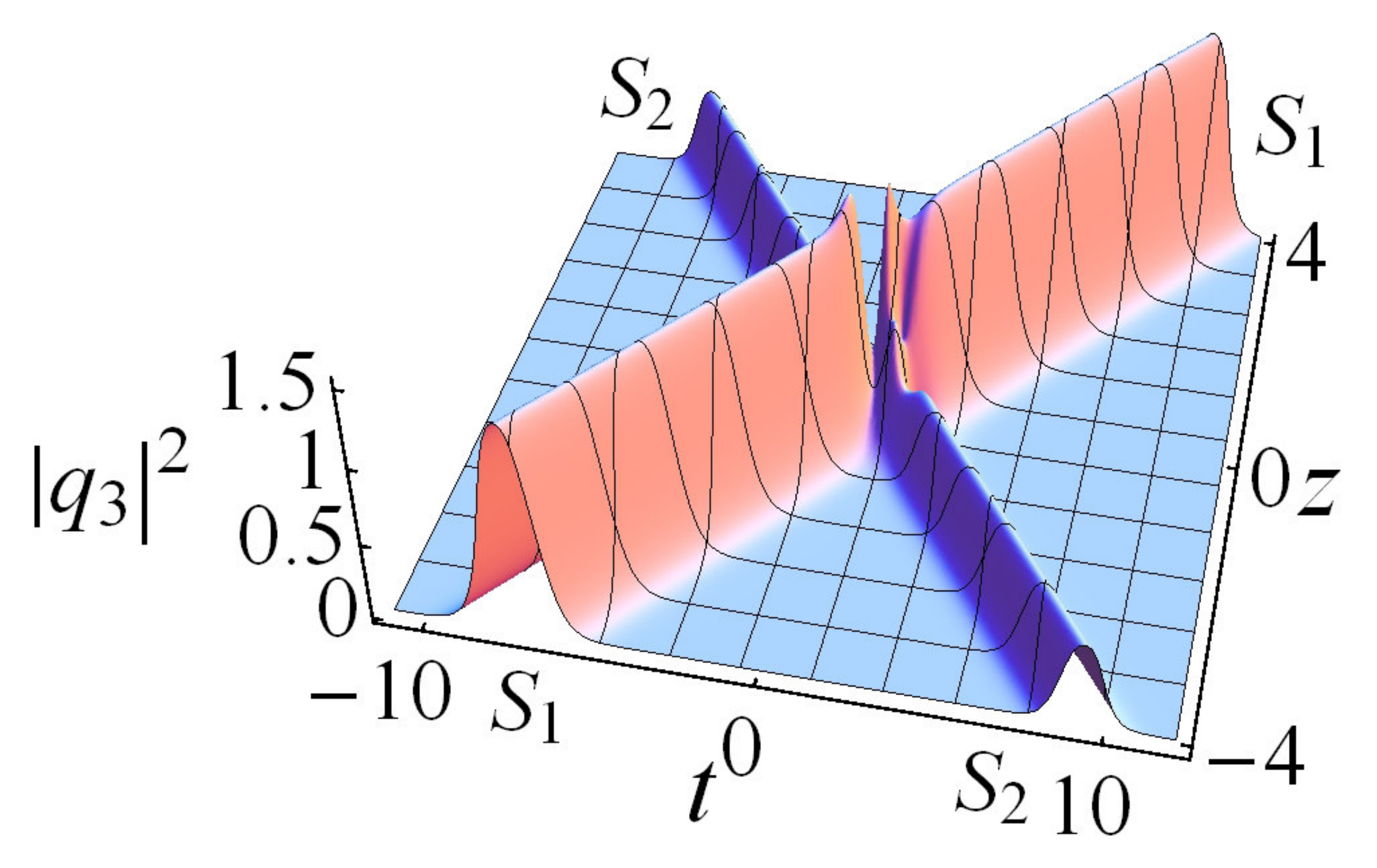}
\caption{Elastic collision of two coherently coupled solitons in $3$-CCNLS system.}
\label{3c-ccsccs}
\end{figure}

\subsection*{(iii) Collision of two ICSs: Type-IV energy sharing collision}
Based on the choice for the ICS in the one-soliton solution, here we obtain two ICSs for the choice $\sum_{j=1}^m (\alpha_u^{(j)})^2=0,~u=1,2$. When two such ICSs collide with each other we get exciting energy sharing collision in the $m$-CCNLS system (\ref{cceqn}). From the asymptotic analysis of (\ref{two-sol-ccnls-a}), the relation between the amplitudes of ICSs before and after collision is obtained as
\bes\bea
A_j^{u+}=T_j^{(u)}~A_j^{u-},  \quad u=1,2, \quad j=1,2,3,...,m. 
\eea
Transition amplitude $T_j^{(u)}$ of soliton $S_u$, $u=1,2$, appearing in the above equation can be written as
 \bea
\hspace{-1.5cm}T_j^{(1)}&=& \frac{\left(1-\hat\lambda_1 +\frac{\alpha_2^{(j)*} \Omega}{\alpha_1^{(j)}~\kappa_{22}}\right)} {\sqrt{1-\hat\lambda_1~\hat\lambda_2+\frac{|\Omega|^2}{\kappa_{11}\kappa_{22}}}} \left(\frac{(k_1-k_2)(k_1^*+k_2)}{(k_1^*-k_2^*)(k_1+k_2^*)}\right)^{\frac{1}{2}}, ~~ j=1,2,3,...,m,\\
\hspace{-1.5cm}T_j^{(2)}&=& - \frac{\sqrt{1-\hat\lambda_1~\hat\lambda_2+\frac{|\Omega|^2}{\kappa_{11}\kappa_{22}}}} {\left(1-\hat\lambda_2+\frac{\alpha_1^{(j)*}\Omega}{\alpha_2^{(j)}~\kappa_{11}}\right)} \left(\frac{(k_1^*-k_2^*)(k_1^*+k_2)}{(k_1-k_2)(k_1+k_2^*)}\right)^{\frac{1}{2}}, ~~ j=1,2,3,...,m,
\eea  \label{tra3c}\ees
where $\Omega=\frac{\gamma }{(k_1-k_2)}\displaystyle\sum_{l=1}^m (\alpha_1^{(l)}\alpha_2^{(l)})$,  $\kappa_{uv}=\frac{\gamma}{(k_u+k_v^*)}\ds\sum_{j=1}^m(\alpha_u^{(j)} \alpha_v^{(j)*})$, $u,v=1,2$, \\$\hat\lambda_1=\frac{\alpha_2^{(j)} \kappa_{12}}{\alpha_1^{(j)} \kappa_{22}}$ and $\hat\lambda_2=\frac{\alpha_1^{(j)} \kappa_{21}}{\alpha_2^{(j)} \kappa_{11}}$.

In general, the ICSs undergo energy sharing collision which involves a change in their amplitudes after collision ($|{T_j}^{(u)}|^2\neq 1$, $u=1,2,~j=1,2,3,...,m$). However, for special choice of $\alpha_1^{(1)}$ parameters $\left(\frac{\alpha_1^{(1)}}{\alpha_2^{(1)}}=\frac{\alpha_1^{(2)}}{\alpha_2^{(2)}}=\frac{\alpha_1^{(3)}}{\alpha_2^{(3)}}=...=\frac{\alpha_1^{(m)}}{\alpha_2^{(m)}}\right)$, we can also get elastic collisions as this choice will lead to $|T_j^{(1)}|^2=|T_j^{(2)}|^2=1$, $j=1,2,3,...,m$. Also, the colliding ICSs $S_u$, $u=1,2$, exhibit phase shift $\Phi_1= \frac{1}{2k_{1R}}\ln \left[\frac{|k_1-k_2|^2}{|k_1+k_2^*|^2} \left(1-\hat\lambda_1~\hat\lambda_2+\frac{|\Omega|^2}{\kappa_{11}\kappa_{22}}\right)\right]$ and $\Phi_2=-\frac{k_{1R}}{k_{2R}}\Phi_1$, respectively which lead to a change in the relative separation distance between the two ICSs, $\Delta t_{12}= \left(1+\frac{k_{1R}}{k_{2R}}\right) \Phi_1$.
\begin{figure}[h]
\centering\includegraphics[width=0.33\linewidth]{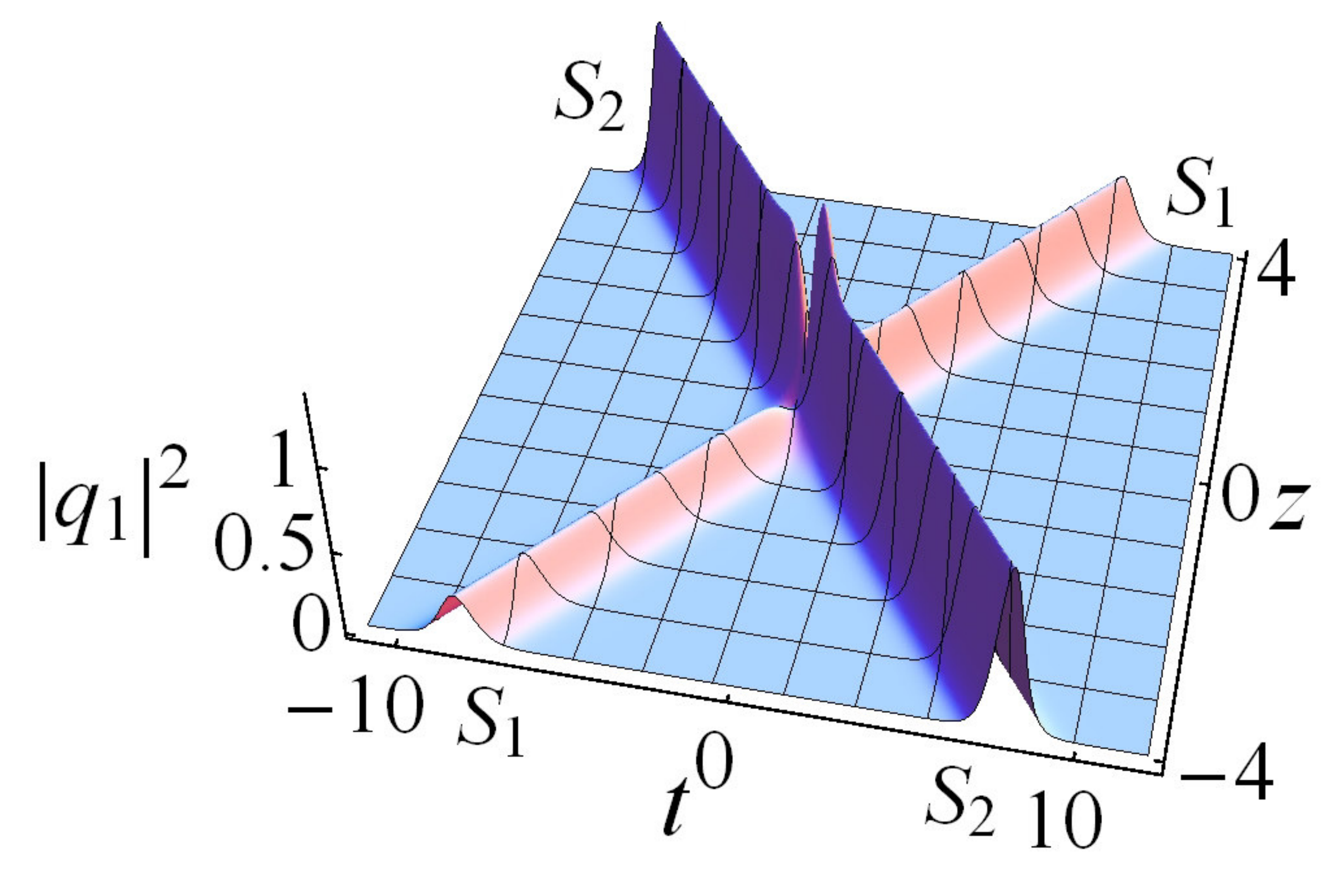}~~~~\includegraphics[width=0.33\linewidth]{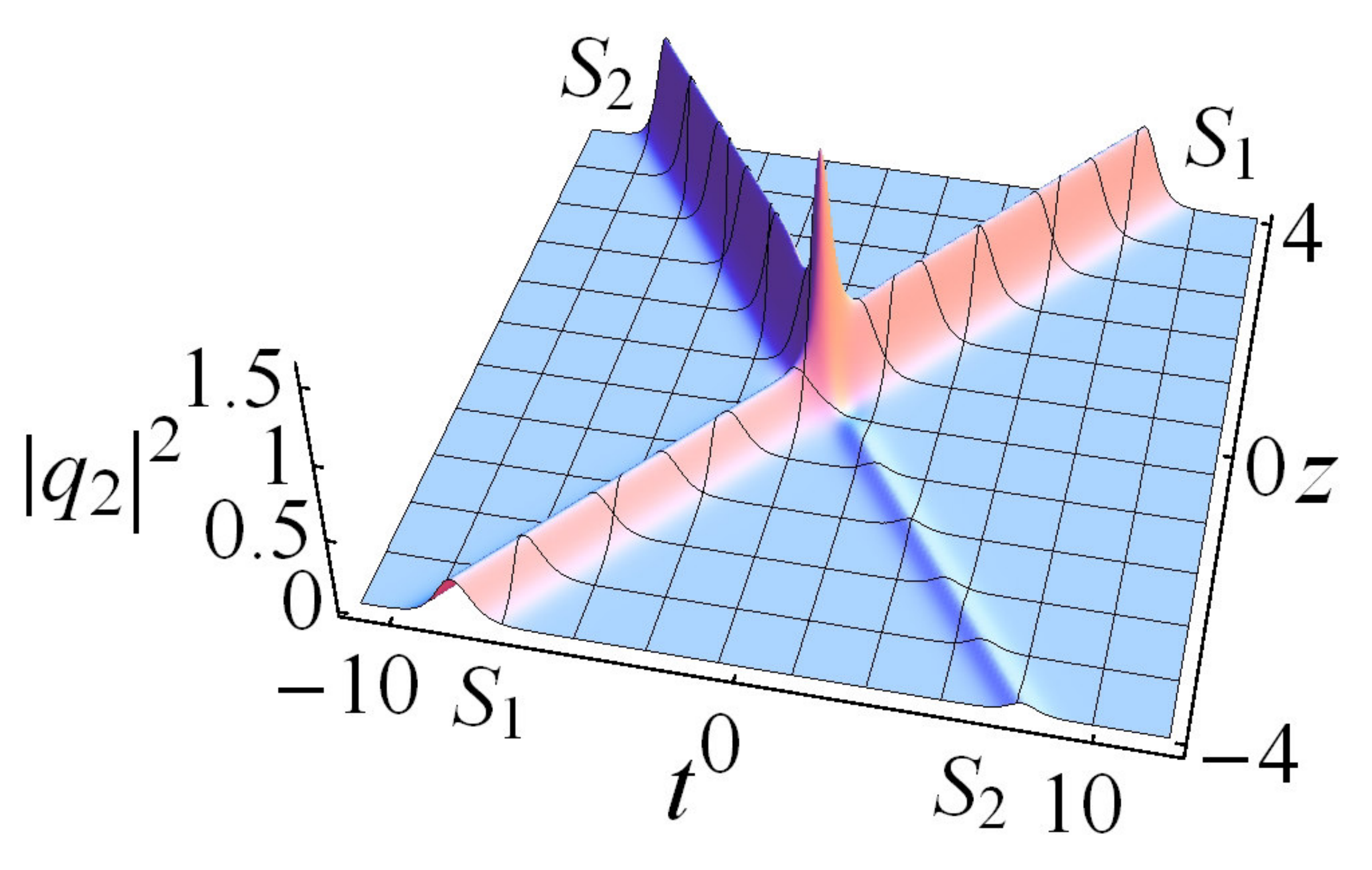}~~~~\includegraphics[width=0.33\linewidth]{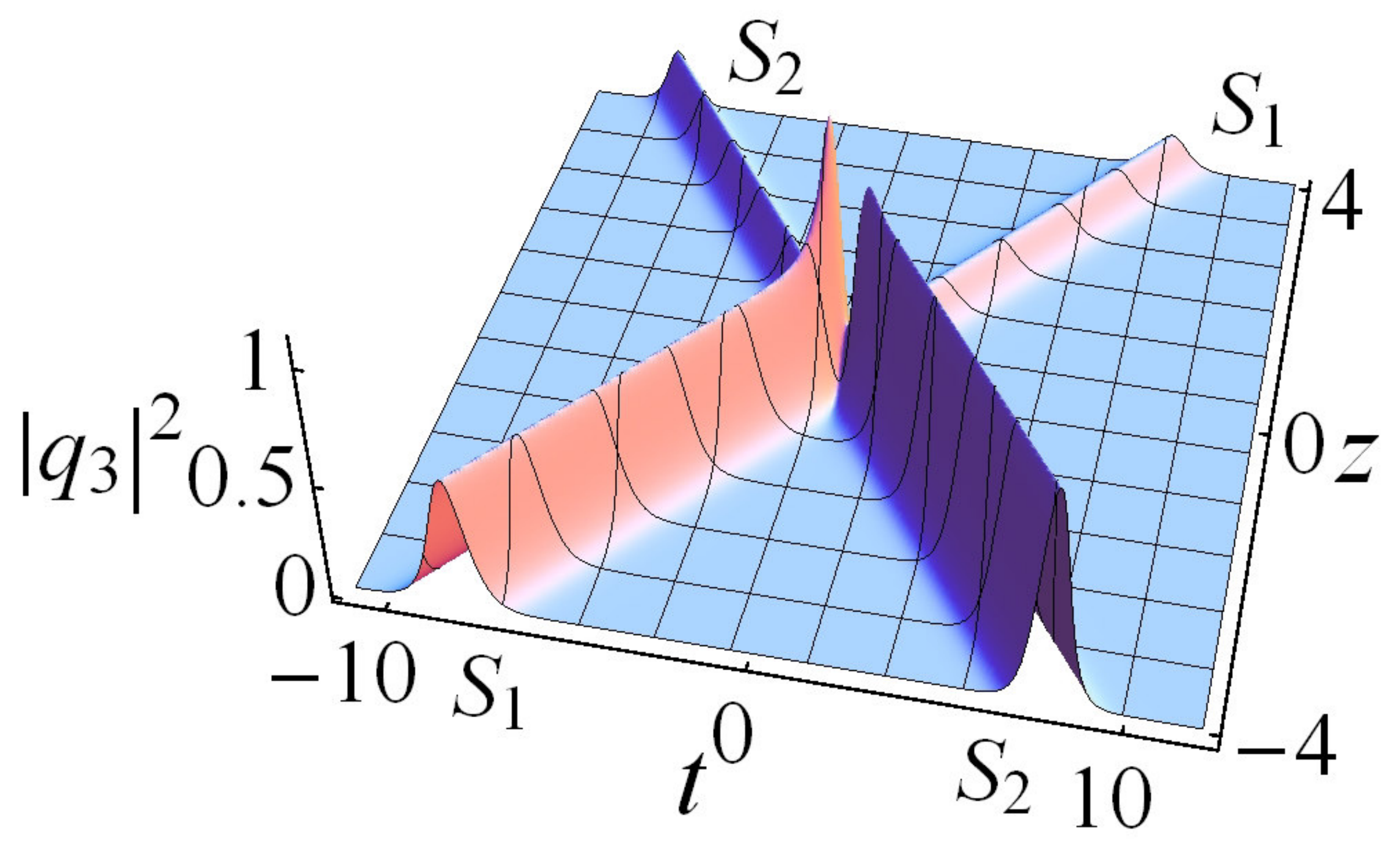}
\caption{Novel type-IV energy sharing collision of two ICSs in $3$-CCNLS system.}
\label{3cc3}
\end{figure}

To facilitate the understanding of ICS-ICS collision first we discuss their collision in the 2-CCNLS system. Here we find that the ICSs always undergo elastic collision. It can be verified that for the 2-CCNLS system the transition amplitudes (intensities) given by Eqn. (\ref{tra3c}) are always uni-modular. However, in the 3-CCNLS system, the ICSs show energy sharing collision in which the energy is not conserved in each $q_j$ component but the total energy among all the components is conserved. We have shown such energy sharing collision of two ICSs in Fig. \ref{3cc3} for the choice $k_1=1.5+i, ~k_2=2-i, ~\gamma=2,~ \alpha_1^{(1)}=\sqrt{2},~ \alpha_1^{(2)}=\sqrt{2},~\alpha_1^{(3)}=2i,~ \alpha_2^{(1)}=\sqrt{8}~i$, $\alpha_2^{(2)}=\sqrt{6}$ and $\alpha_2^{(3)}=\sqrt{2}$. It is interesting to note that only CCS undergoes intensity change during its collision with ICS (i.e. CCS-ICS collision) and ICS remains unaltered. But in the case of collision between two ICSs, the intensity gets altered in both solitons. In Fig. \ref{3cc3}, the two colliding solitons always undergo an enhancement in their intensities in $q_1$ and $q_2$ components while they experience suppression in the $q_3$ component. This type of collision between two ICSs takes place only in the $m$-CCNLS system with $m\geq 3$ and one can not expect such energy sharing collision in the 2-CCNLS system, which exhibits elastic collision always. We refer to this collision scenario in which the solitons undergo same kind of switching in a given given component with commensurate changes in the other components as Type-IV energy sharing collision. In addition to this type-IV energy sharing collision, the elastic collision of two ICSs can also result for the choice $\frac{\alpha_1^{(1)}}{\alpha_2^{(1)}}=\frac{\alpha_1^{(2)}}{\alpha_2^{(2)}}=\frac{\alpha_1^{(3)}}{\alpha_2^{(3)}}$.

\section{Conclusion}
We have investigated various types of soliton collisions in multicomponent nonlinear Schr\"odinger type systems with different nonlinearities. We have revealed type-I and type-II energy sharing collisions in the two-component Manakov and mixed ICNLS systems, respectively, where the solitons undergo opposite and same kind of energy sharing nature among the two components. Such energy sharing collisions can also be observed in their multicomponent counterparts with $m\geq 3$. Then we have discussed the interesting energy switching collision scenario in coherently coupled nonlinear Schr\"odinger system which admits various soliton profiles like single-hump, double-hump and flat-top structures. This energy switching collision does not conserve the energy in a given component instead the total energy of all the components is conserved. Particularly, we have demonstrated that during its collision ICS, the CCS undergoes energy switching leaving the ICS unaltered. We have referred this collision as type-III energy sharing collision. Also, another interesting type-IV energy sharing collision has been identified in the $m$-CCNLS system with $m \geq 3$. Additionally, elastic collision processes of different bright-solitons are observed for special choices of soliton parameters. The reported four types of energy sharing collisions of bright solitons will find applications in the context of soliton collision based optical computing, optical switching devices, etc.

\section*{Acknowledgments}
TK and KS thank the organizers of the National Mathematics Initiative workshop on Nonlinear Integrable Systems and their Applications 2014. TK and KS also thank the principal and management of Bishop Heber College for constant support and encouragement. KS is grateful to the support of Council of Scientific and Industrial Research, Govt. of India, with a Senior Research Fellowship. MV acknowledges the financial support from UGC-Dr. D. S. Kothari post-doctoral fellowship.

\appendix \section{Bright one-soliton solution of the $m$-CCNLS system (\ref{cceqn})} \label{one-sol-ccnls}
The bright one-soliton solution of the $m$-CCNLS equation (\ref{cceqn}) obtained by using the non-standard approach of Hirota's bilinearization method \cite{Kanna2011jpa} can be written as
\bes\bea
q_j&=&\frac{\alpha_1^{(j)} e^{\eta_1}+e^{2\eta_1+\eta_1^*+\delta_{11}^{(j)}}}{1+e^{\eta_1+\eta_1^*+R_1}+e^{2\eta_1+2\eta_1^*+\epsilon_{11}}},\quad ~j=1,2,3,...,m, \label{os3cc1}
\eea
\noindent where 
\bea e^{\delta_{11}^{(j)}}=\frac{\gamma \alpha_1^{(j)*} \Gamma_1}{2 (k_1+k_1^*)^2}, \quad 
e^{R_1}=\frac{\kappa_{11}}{(k_1+k_1^*)},\quad 
e^{\epsilon_{11}}=\frac{\gamma^2 |\Gamma_1|^2}{4 (k_1+k_1^*)^4},
\eea
in which 
\bea
\Gamma_1=\sum_{j=1}^m (\alpha_1^{(j)})^2, \qquad \kappa_{11}=\frac{\gamma \sum_{j=1}^m {|\alpha_1^{(j)}|^2}}{(k_1+k_1^*)}.
\eea\ees
Here the auxiliary function $s$ is obtained as $s=\sum_{j=1}^m (\alpha_1^{(j)})^2 e^{2\eta_1}$.

\section{Bright two-soliton solution of the $m$-CCNLS system (\ref{cceqn})}\label{two-sol-ccnls}
The bright two-soliton solution of system (\ref{cceqn}) can be written as \cite{Kanna2011jpa}
\bes\label{two-sol-ccnls-a}
\bea
q_j=\frac{g^{(j)}}{f},\quad j=1,2,3,...,m, 
\eea
where
\bea
g^{(j)}&=&\sum (\alpha _u^{(j)} e^{\eta _u}) +\sum  (e^{2 \eta _u+\eta _v^{*}+\delta_{uv}^{(j)}})\nonumber\\
&&+\sum  (e^{\eta_1+\eta_2+\eta_u^*+\delta_u^{(j)}})+\sum (e^{2\eta_u+2\eta_v^*+\eta_{3-u}+\mu_{uv}^{(j)}})\nonumber \\
&&+e^{\eta_1+\eta_1^*+\eta_2+\eta_2^*}\left(\sum e^{\eta_u+\mu_u^{(j)}}+\sum e^{\eta_1+\eta_2+\eta_u^*+\phi_u^{(j)}}\right),\nonumber \\
&&\qquad\qquad\qquad\qquad\qquad\qquad\qquad j=1,2,3,...,m,\\
f&=&1+\sum (e^{\eta_u+\eta_u^*+R_u})+e^{\eta_1+\eta_2^*+\delta_0}+e^{\eta_2+\eta_1^*+\delta_0^*}\nonumber\\
&&+\sum (e^{2\eta_u+2\eta_v^*+\epsilon_{uv}})+e^{\eta_1^*+\eta_2^*}\sum (e^{2\eta_u+\tau_u})+e^{\eta_1+\eta_2}\sum (e^{2\eta_u^*+\tau_u^*})\nonumber\\
&&+e^{\eta_1+\eta_1^*+\eta_2+\eta_2^*}\left(e^{R_3}+\sum e^{\eta_u+\eta_v^*+\theta_{uv}}+e^{\eta_1+\eta_1^*+\eta_2+\eta_2^*+R_4}\right),~~
\eea
and the auxiliary function $s$ takes the form
\bea
s&=&\sum (\Gamma_u e^{2 \eta _u})+\Gamma_3 e^{\eta _1+\eta _2} +\sum (e^{\eta _u+2\eta _{3-u}+\eta _v^*+\lambda _{uv}}) \nonumber\\
&&\qquad + e^{2 \eta _1+2 \eta _2} \left(\sum e^{2\eta _u^*+\lambda _u}+e^{\eta_1^*+\eta _2^*+\lambda _3}\right).
\eea \ees
Here $\eta_u=k_u(t+ik_u z),~u=1,2$, the summation is taken over $u$ and $v$ for $u,v=1,2$, and the expressions for various other quantities are given below.
\bea
e^{R_u}&=&\frac{\kappa _{uu}}{(k_u+k_u^*)},~~ e^{\delta _0}=\frac{ \kappa _{12}}{(k_1+k_2^*)},~~
 e^{\delta _0^*}=\frac{ \kappa _{21}}{(k_2+k_1^*)},\nonumber\\
e^{\delta _{uv}^{(j)}}&=&\frac{\gamma \alpha_v^{(j)*}\Gamma_u}{2 (k_u+k_v^*)^2},\quad 
e^{\delta _u^{(j)}}=\frac{\gamma \alpha _u^{(j)*} \Gamma_3+(k_1-k_2) (\alpha _1^{(j)} \kappa _{2u}-\alpha _2^{(j)} \kappa_{1u})}{(k_1+k_u^*) (k_2+k_u^*)},\nonumber\\
e^{\epsilon _{uv}}&=&\frac{ \gamma ^2 \Gamma_u \Gamma_v^*}{4 (k_u+k_v^*)^4},~\quad e^{\tau _u}=\frac{\gamma ^2 \Gamma_3^* \Gamma_u}{2 (k_u+k_1^*)^2 (k_u+k_2^*)^2},\nonumber\\ 
e^{\lambda _{uv}}&=&\frac{(k_1-k_2)^2 \kappa_{uv} \Gamma_{3-u}}{(k_u+k_v^*)(k_{3-u}+k_v^*)^2},\quad 
e^{\mu _{uv}^{(j)}}=\frac{\gamma^2 (k_1-k_2)^2 \alpha _{3-u}^{(j)} \Gamma_u \Gamma_v^*}{4 (k_u+k_v^*)^4 (k_{3-u}+k_v^*)^2},\nonumber\\
e^{\theta _{uv}}&=&\frac{\gamma ^2|k_1-k_2|^4}{{4\tilde{D}}(k_u+k_v^*)^2} \Gamma_u \Gamma_v^* \kappa_{3-u~3-v},\quad
e^{\lambda _u}=\frac{\gamma ^2(k_1-k_2)^4 \Gamma_1 \Gamma_2 \Gamma_u^*}{4 (k_1+k_u^*)^4 (k_2+k_u^*)^4},\nonumber\\
e^{\lambda _3}&=&\frac{\gamma^2(k_1-k_2)^4}{2\tilde{D}} \Gamma_1 \Gamma_2 \Gamma_3^*,\quad
e^{\phi_u^{(j)}}=\frac{\gamma ^3 (k_1-k_2)^4 (k_1^*-k_2^*)^2} {8\tilde{D} {(k_1+k_u^*)^2(k_2+k_u^*)^2}} {\alpha _{3-u}^{(j)*} \Gamma_1 \Gamma_2 \Gamma_u^*},\nonumber\\
e^{R_3}&=& \frac{|k_1-k_2|^2(\kappa_{11}\kappa_{22}-\kappa_{12}\kappa_{21})+\gamma^2 |\Gamma_3|^2}{(k_1+k_1^*)|k_1+k_2^*|^2(k_2+k_2^*)},\quad 
e^{R_4}=\frac{\gamma ^4 |k_1-k_2|^8}{16\tilde{D}^2} \Gamma_1 \Gamma_2 \Gamma_1^* \Gamma_2^*,\nonumber\\
e^{\mu_u^{(j)}}&=&\frac{(k_1-k_2)^2 \gamma}{2\tilde{D}}{\Gamma_u} (k_{3-u}+k_1^*)(k_{3-u}+k_2^*)\nonumber\\
&&\quad \times \left[\gamma \alpha_{3-u}^{(j)}\Gamma_3^* + (k_1^*-k_2^*) (\alpha_1^{(j)*}\kappa_{3-u 2}-\alpha_2^{(j)*}\kappa_{3-u 1})\right],\nonumber
\eea
where
\bea
\tilde{D}&=&(k_1+k_1^*)^2(k_1^*+k_2)^2 (k_1+k_2^*)^2 (k_2+k_2^*)^2,\nonumber\\ \kappa_{uv}&=&\frac{\gamma}{(k_u+k_v^*)}\ds\sum_{j=1}^m(\alpha_u^{(j)} \alpha_v^{(j)*}),\nonumber\\ 
\Gamma_1&=&\ds\sum_{j=1}^m (\alpha_1^{(j)})^2,\quad \Gamma_2=\ds\sum_{j=1}^m (\alpha_2^{(j)})^2,\quad \Gamma_3=\ds\sum_{j=1}^m (\alpha_1^{(j)}\alpha_2^{(j)}).\nonumber
\eea
In the above expressions $u,v=1,2$ and $j=1,2,3,...,m$.

\end{document}